\documentclass[iop]{emulateapj}
\usepackage{hyperref}
\usepackage{amsmath}
\usepackage[flushleft]{threeparttable}

\shorttitle{Core Emergence in Infrared Dark Clouds}
\shortauthors{Kong et al.}

\begin{document}

\title{Core Emergence in a Massive Infrared Dark Cloud:\\A Comparison Between Mid-IR Extinction and 1.3 mm Emission}

\author{Shuo Kong\altaffilmark{1}}
\affil{Dept. of Astronomy, Yale University, New Haven, Connecticut 06511, USA}

\author{Jonathan C. Tan\altaffilmark{2,3}}
\affil{Dept. of Space, Earth and Environment, Chalmers University of Technology, Gothenburg, Sweden}
\affil{Dept. of Astronomy, University of Virginia, Charlottesville, Virginia 22904, USA}

\author{H\'ector G. Arce\altaffilmark{1}}
\affil{Dept. of Astronomy, Yale University, New Haven, Connecticut 06511, USA}

\author{Paola Caselli\altaffilmark{4}}
\affil{Max-Planck-Institute for Extraterrestrial Physics (MPE), Giessenbachstr. 1, D-85748 Garching, Germany}

\author{Francesco Fontani\altaffilmark{5}}
\affil{INAF - Osservatorio Astrofisico di Arcetri, I-50125, Florence, Italy}

\author{Michael J. Butler\altaffilmark{6}}
\affil{Max Planck Institute for Astronomy, K\"onigstuhl 17, 69117 Heidelberg, Germany}

\begin{abstract}
Stars are born from dense cores in molecular clouds.  Observationally,
it is crucial to capture the formation of cores in order to understand
the necessary conditions and rate of the star formation process. The
{\it Atacama Large Mm/sub-mm Array} (ALMA) is extremely powerful for
identifying dense gas structures, including cores, at mm wavelengths
via their dust continuum emission. Here we use ALMA to carry out a
survey of dense gas and cores in the central region of the massive
($\sim10^5\:M_\odot$) Infrared Dark Cloud (IRDC) G28.37+0.07.  The
observation consists of a mosaic of 86 pointings of the 12m-array and
produces an unprecedented view of the densest structures of this
IRDC. In this first paper about this data set, we focus on a
comparison between the 1.3 mm continuum emission and a mid-infrared
(MIR) extinction map of the IRDC. This allows estimation of the
``dense gas'' detection probability function (DPF), i.e., as a
function of the local mass surface density, $\Sigma$, for various
choices of thresholds of mm continuum emission to define ``dense
gas''. We then estimate the dense gas mass fraction, $f_{\rm dg}$, in
the central region of the IRDC and, via extrapolation with the DPF and
the known $\Sigma$ probability distribution function, to the
larger-scale surrounding regions, finding values of about 5\% to 15\%
for the fiducial choice of threshold. We argue that this observed dense
gas is a good tracer of the protostellar core population and, in this context,
estimate a star formation efficiency per free-fall time in the central
IRDC region of $\epsilon_{\rm ff}\sim$10\%, with approximately a factor
of two systematic uncertainties.
\end{abstract}


\section{Introduction}

Dense cores, as the birthplace of stars, are the focus of intense
theoretical and observational study, in particular for understanding
the initial conditions and efficiency of star formation \citep[see,
  e.g.,][]{2007ARA&A..45..339B,2014prpl.conf..149T,2014prpl.conf...53O}.
One theory of core formation is that of gravito-turbulent
fragmentation where dense, gravitationally unstable cores are created
in density perturbations arising from compressions in supersonically
turbulent molecular gas
\citep[e.g.,][]{2002ApJ...576..870P,2005ApJ...630..250K,
  2008ApJ...684..395H,2014ApJ...796...75C}.  In the theory of
\citet{2005ApJ...630..250K} \citep[see also][]{2011ApJ...741L..22P}
the rate of star formation is linked to the Mach number and virial
parameter (i.e., degree of gravitational boundedness) of the cloud,
although it should be remembered that most of the numerical simulation
results that underpin these models are of periodic box turbulence
(rather than of global cloud structures) and have relatively weak
magnetic fields (i.e., are globally magnetically supercritical)
\citep[see][]{2016IAUS..315..154T}.  Models of core formation mediated
by magnetic fields, e.g., via ambipolar diffusion
\citep{2009MNRAS.399L..94K,2017ApJ...848...50C}, have also been
proposed.

Giant molecular clouds (GMCs) in the Milky Way are known to have a low
star formation efficiency (SFE)
\citep{1974ApJ...192L.149Z,2012ARA&A..50..531K}, where this quantity
is most naturally evaluated as the fraction of mass that forms stars
in one local free-fall time of the gas, i.e., $\epsilon_{\rm ff}$,
which is seen to have a value of about 0.02.
\citet{2007ApJ...654..304K} extended the analysis methods of
\citet{1974ApJ...192L.149Z} to denser gas structures, such as Infrared
Dark Clouds (IRDCs) and embedded clusters, finding similar values of
$\epsilon_{\rm ff}$. \citet{2011ApJ...729..133M} and
\citet{2016ApJ...833..229L} have pointed out there is a large
dispersion in $\epsilon_{\rm ff}$ in Galactic GMCs, but the average
value in the population is still low and consistent with prior
estimates. Theoretical models of star formation rates (SFRs)
regulated by turbulence
\citep[e.g.,][]{2005ApJ...630..250K,2011ApJ...743L..29H,2012ApJ...754...71K}
are one way to explain the low averaged observed values of
$\epsilon_{\rm ff}$. However, \citet{2016ApJ...833..229L} have noted
that the high dispersion in $\epsilon_{\rm ff}$ is not accounted for
in these models. This may imply a role for more stochastic processes,
such as triggering by collisions of magnetically supported GMCs
inducing bursts of star formation activity
\citep{1986ApJ...310L..77S,2000ApJ...536..173T,
  2015ApJ...811...56W,2017ApJ...841...88W}.

In order to observationally capture the formation of cores 
and have a sneak peek of the signature of low SFE,
we can compare features that are specifically sensitive to
the core formation with features representing the host cloud. 
Dust continuum emission at mm wavelength
is sensitive to star-forming cores and often used as a core tracer
\citep[e.g.,][]{2007ARA&A..45..339B,2017ApJ...841...97S}.
If observed with interferometers like ALMA, 
mm continuum emission can potentially pinpoint dense, star-forming
cores embedded in a molecular cloud because the extended emission
is resolved out by interferometers. Moreover, continuum emission
does not depend on chemical abundance variations (such as
chemical depletion) which can affect molecular-line-defined cores. 
The host cloud, and its density structure, can be
traced by far-infrared dust emission \citep{2014prpl.conf...27A}
or dust extinction \citep{2009A&A...493..735L,2009ApJ...696..484B}.
The latter has been used to construct $\Sigma$ probability distribution 
functions \citep[$\Sigma$-PDF, e.g.,][]{2009A&A...508L..35K,
2014ApJ...782L..30B,2015A&A...577L...6S,2016ApJ...829L..19L}.
Such distributions appear to have a log-normal component, perhaps arising
from turbulence, together with a high-$\Sigma$ power law tail, perhaps
due to self-gravity. However, observationally it is
challenging to accurately measure the $\Sigma$-PDF 
\citep[e.g.,][]{2017A&A...606L...2A,2017arXiv170709356C}.
Numerical simulations of
molecular clouds also aim to reproduce the $\Sigma$-PDF and understand
its dependence on the included physics \citep[see,
  e.g.,][]{2011ApJ...731...59C,2013ApJ...763...51F,2015ApJ...806..226M}.

In this paper, we study the relationship between mm continuum emitting
dense gas structures and the local mass surface density of their
hosting cloud to better understand the conditions of core
formation. Our target cloud is IRDC G28.37+0.07 (also referred to as
IRDC C) from the sample of \citet[][hereafter BT09,
  BT12]{2009ApJ...696..484B,2012ApJ...754....5B} at an estimated
kinematic distance of 5~kpc. Specifically, we compare the mid-infrared
(MIR) extinction map of the IRDC \citep[][hereafter
  BTK14]{2014ApJ...782L..30B} with a newly acquired 1.3 mm continuum
image, observed with ALMA.

\section{Data}\label{sec:obs}

\subsection{ALMA Observations}\label{sec:obsalma}

The observations were carried out between 24-Jun-2016 and 30-Jun-2016
(UTC) ({\it ALMA} Cycle 3), under the project 2015.1.00183.S (PI:
Kong).  Forty-six 12-m antennas were used during the observation in
C40-4 configuration.  A custom mosaic with 86 pointings was used to
cover the majority of the central dark regions of IRDC G28.37+0.07.
The observations were in band 6 ($\sim$ 231 GHz).  A baseband of 1.8
GHz was used for the 1.3~mm continuum observation (the effective
bandwidth for continuum imaging is $\sim1.4\:$GHz due to the exclusion
of the $^{12}$CO(2-1) molecular line).  Three other basebands were set
for molecular line observations.  In this paper, we focus on the
continuum data; we defer the analysis of the molecular line data to a
future paper.

The observations consist of 6 scheduling blocks, each having roughly
50 minutes on-source integration time.  J1751+0939 and J1924-2914 were
used as bandpass calibrators.  J1751+0939, J1924-2914, and Titan were
used as flux calibrators.  J1851+0035 was used as the phase
calibrator.  The typical system temperature was 80~K.  The mosaic
image was cleaned using the standard {\it clean} task in CASA. Briggs
weighting with a robust number of 0.5 was applied.  No
self-calibration was done.  In order to perform a pixel-by-pixel
comparison with the MIREX image (see following section), we applied an
outer uv-taper to match the ALMA synthesized beam to the MIREX beam (2\arcsec).  
A pixel scale of 0.4\arcsec~was adopted in the {\it clean} task.  Then
we re-binned the images to have 1.2\arcsec\ pixels to match the MIREX
pixel scale.  The resulting sensitivity at map center is $\sigma_{\rm center}=$
0.2 mJy per 2\arcsec~beam.  The maximum recoverable scale of the ALMA
continuum image is $\sim$20\arcsec\ (corresponding to the 
shortest baseline of 10 k$\lambda$ with $\lambda$ being 1.3 mm).

\subsection{The MIREX image}\label{sec:mirex}

The MIR extinction (MIREX) map of IRDC G28.37+0.07 was first developed
by BT09 and BT12 using Spitzer 8~$\rm \mu m$ GLIMPSE imaging data
\citep{2009PASP..121..213C}. It was merged with a lower resolution NIR
extinction map by \citet{2013A&A...549A..53K}, which improves accuracy
at lower values of $\Sigma$. Finally, the map was refined by
\citet{2014ApJ...782L..30B} by using an analysis of deeper archival
      {\it Spitzer}-IRAC imaging,
which enables the highest dynamic range of $\Sigma$ to be probed. In
general, the method of MIREX mapping involves estimating the intensity
of the diffuse background emission, i.e., from the diffuse Galactic
ISM, via interpolation from surrounding regions, and estimating,
empirically, the level of the foreground emission. Then, given an
estimate of the dust opacity at 8~$\rm \mu m$ (averaged over the
Spitzer IRAC Band 4) and a dust to gas mass ratio, the total mass
surface density can be calculated by solving the simple 1D radiative
transfer equation, given the observed intensities emerging from the
cloud.  The spatial resolution achieved in the map is 2\arcsec with a
pixel scale of 1.2\arcsec, set by the resolution of the {\it
  Spitzer}-IRAC data.

There are several effects that lead to systematic errors in the MIREX
maps. One problem is that in regions containing local bright MIR
source, the extinction is contaminated by the source.
Another problem is that in some regions the IRDCs become
very optically thick, so only a lower limit on $\Sigma$ can be
estimated. These regions are referred to as being ``saturated'' in the
MIREX map (their presence allows the measurement of the diffuse
foreground emission, assumed to be spatially constant). Local
fluctuations in the background will lead to errors, since it is
modeled as a smoothly varying source. Zero point offsets of up to
$\sim0.1\:{\rm g\:cm}^{-2}$ are present, which are partially corrected
for by calibration with NIR extinction maps
\citep{2013A&A...549A..53K}. Still, the zero point uncertainty is
present at a level estimated to be $\sim0.02\:{\rm g\:cm}^{-2}$ (i.e.,
$A_V\sim 4$~mag or so).

\section{Results}\label{sec:results}

\subsection{Comparison between 1.3~mm Dust Continuum Emission and MIR Extinction}\label{sec:compare}

\begin{figure*}[htb!]
\epsscale{1.15}
\plotone{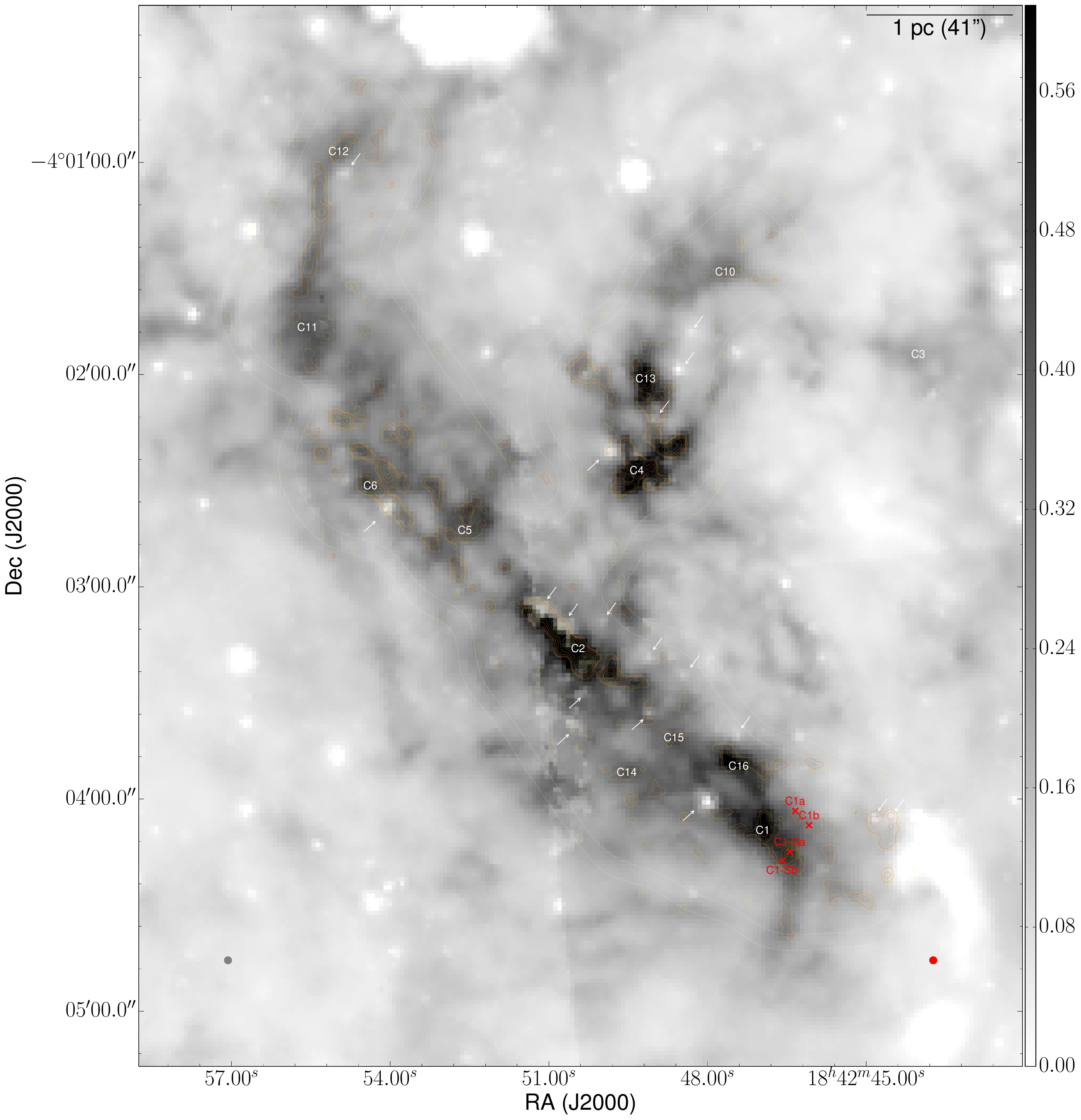}
\caption{
Greyscale: MIREX mass surface density map from BTK14 (scale in g
cm$^{-2}$). The angular resolution of the map is shown as the gray
filled circle at lower-left. ``C1, C2, C3...'' label extinction peaks
from BTK14. The white arrows point to possible embedded protostars
that show as local enhancements in the $\rm 8\:\mu m$ image, which
produce local ``holes'' in the MIREX map.  Contours: ALMA 1.3~mm
continuum mosaic. The contours range from SNR = 2, 3, 5, 10,
20, 40, 60, ... with the rms noise at map center $\sigma_{\rm center} \sim
0.2 \rm mJy\:beam^{-1}$. The two red contours highlight SNR = 3 and 10.
The synthesized beam is shown as the red
filled ellipse at lower-right.  The two white enclosing contours show
primary-beam responses of 0.3 (outer) and 0.5 (inner).
\label{fig:cont_mirex}}
\end{figure*}

Figure \ref{fig:cont_mirex} shows the comparison between the cloud
mass surface density from the MIREX map and the 1.3 mm dust continuum
emission observed by ALMA. In general, the MIREX image shows mainly
$\Sigma\ga0.2\:$g~cm$^{-2}$ pixels in the ALMA-mosaicked region.  They
correspond to relatively dark regions in the original Spitzer IRAC
8$\rm \mu m$ image. The MIREX map reveals features via dust absorption
(depending on total $\Sigma$), while the ALMA image shows dust
emission (depending on total $\Sigma$ and dust temperature). Another
difference arises due to ALMA filtering out low spatial frequency
(larger-scale) structures.  In our case, the recoverable physical
scales range from 10,000~AU (0.05~pc, 2\arcsec) (after uv-tapering) to
approximately 100,000~AU (0.48~pc, 20\arcsec). 
We note that the Jeans length
\begin{equation}\label{eq:jeanslength}
\lambda_{J} = 0.10\left(\frac{T}{15\:\rm K}\right)^{1/2}\left(\frac{n_{\rm H}}{10^5\: \rm cm^{-3}}\right)^{-1/2}\:\rm pc
\end{equation}
is in the range of recovered scales, given typical conditions of
ambient gas in the IRDC. Consequently, while the extinction map tracks
the total column density, the ALMA continuum image pinpoints compact,
dense and warmer structures, i.e., expected to be protostellar
cores. Thus, through comparison with the extinction map, the ALMA
image shows us where such dense, likely star-forming, structures
emerge from the cloud.

We now give a brief overview of several of the regions seen in the
map. Dense ``cores/clumps'' C1 to C16 were identified in the MIREX map
by BT12 and BTK14. The continuum cores in the south-west C1 region
were studied by
\citet{2013ApJ...779...96T,2016ApJ...821L...3T,2017arXiv170105953K}.
C1-Sa and C1-Sb have been identified as protostellar cores and C1a
and C1b as candidate protostellar cores. A massive pre-stellar core
candidate, C1-S, identified by $\rm N_2D^+(3-2)$ emission by
\citet{2013ApJ...779...96T}, sits between C1-Sa and C1-Sb, but has
relatively faint 1.3~mm continuum emission. C1 is the location of the
C1-N core, which is another massive pre-stellar core candidate
identified by its $\rm N_2D^+(3-2)$ emission. We note that most of the
protostellar cores (including the relatively low-mass
$\sim2\:M_\odot$ C1-Sb core) and some massive pre-stellar cores are
well-detected in the ALMA continuum image.

Moving to the NE, several other sources are seen in the region,
including the C14, C15 and C16 core/clumps. Next we come to the C2
region, which corresponds to the ``P1 clump'' studied by
\citet[][]{2009ApJ...696..268Z,2015ApJ...804..141Z}.  They identified
a linear chain of five main continuum structures, with a hint of a
sixth core/clump at the SW end.  Here we confirm the detection of this
sixth, weaker continuum structure. Like the other cores, it also
corresponds to a high-$\Sigma$ peak in the MIREX map. With the higher
resolution ($\sim0.7\arcsec$) observations of
\citet{2015ApJ...804..141Z} a few tens of cores were identified in the
C2 region down to sub-solar masses, with many of these seen to be
protostellar by the presence of bipolar CO outflows.

North-west of C2 is a region containing C4, C10 and C13, with most of
the mass concentrated near C4 and C13. Several distinct mm continuum
peaks are visible in this region. Continuing north-east from C2 is the
sequence of MIR dark core/clumps C5 and C6, which contain a cluster of
mm emission cores, then the sparser C11 and C12. Between C11 and C12
there is a narrow filament seen in mm continuum emission, which
closely follows the morphology seen in the MIREX map. This filament
shows signs of fragmenting into several cores (including C12), but may
be at an earlier stage of evolution compared to the more fragmented
regions described above, such as C5/C6, C4/C13 and perhaps C2.


Globally, Figure \ref{fig:cont_mirex} shows that the 1.3~mm continuum
structures follow the extinction features quite well, i.e., they tend
to be found in high-$\Sigma$ regions of the MIREX map. For example, in
the region around C4, the cloud shows very good agreement between the
continuum emission and high-$\Sigma$ pixels. On the other hand, MIREX
high-$\Sigma$ regions do not always show mm continuum emission. This
is illustrated in the region around C11, where it shows few robust 1.3
mm continuum detections. Being in a high-$\Sigma$ region is a
necessary, but not sufficient, condition for the presence of strong mm
continuum cores.

\begin{figure*}[htb!]
\epsscale{0.5}
\plotone{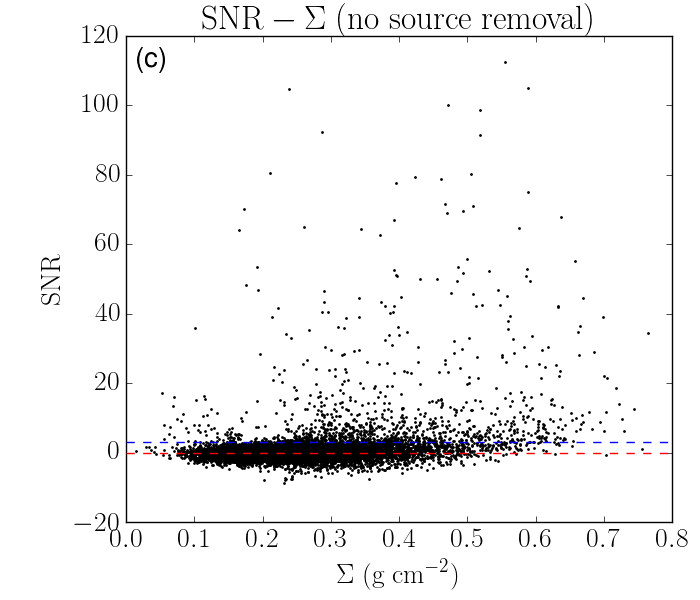}
\plotone{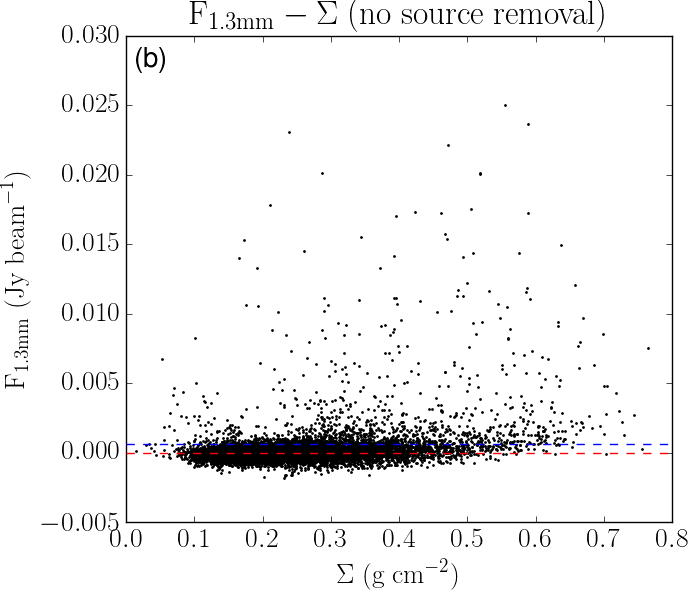}\\
\plotone{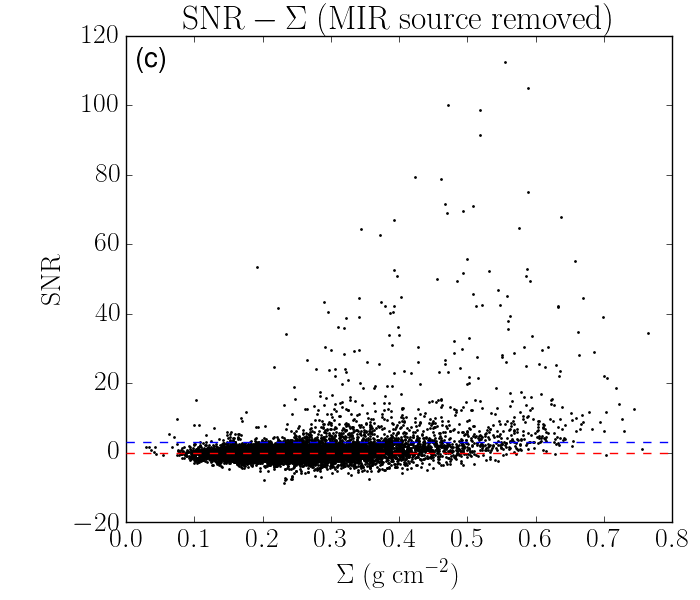}
\plotone{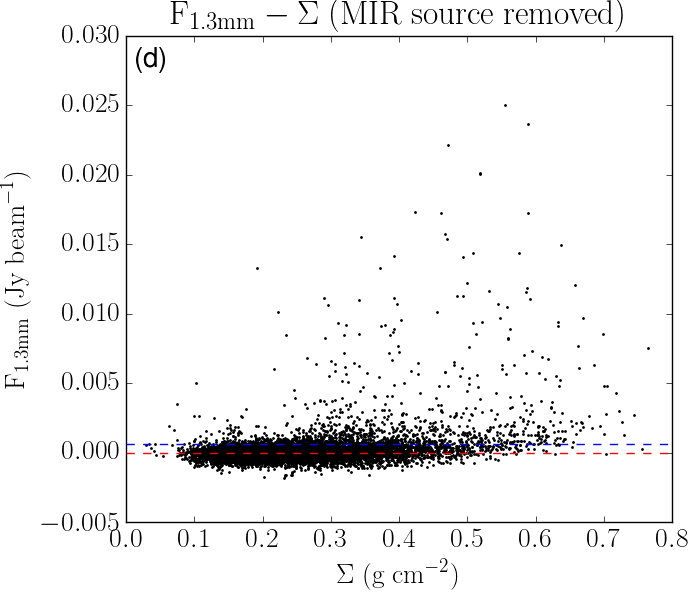}\\
\caption{
{\bf (a)} Pixel-by-pixel comparison between SNR and $\Sigma$.
The red dashed line shows the zero point of the continuum image. 
The blue dashed line shows the SNR = 3 noise level.
The map boundary is defined where the primary-beam response is 0.5.
{\bf (b)} Pixel-by-pixel comparison between $F_{\rm 1.3mm}$ and $\Sigma$.
The map boundary is defined where the primary-beam response is 0.5 
(see Figure \ref{fig:cont_mirex}).
The noise at the map center $\sigma_{\rm center}$ (indicated by the blue dashed line) 
is a factor of 2 smaller than at the edge.
{\bf (c)} Same as (a) but removing the embedded sources.
See \S\ref{subsec:dp}.
{\bf (d)} Same as (b) but removing the embedded sources.
See \S\ref{subsec:dp}.
\label{fig:compareflux}}
\end{figure*}

In order to reveal more quantitatively the large-scale mass surface
density conditions needed for the formation of 1.3~mm continuum
emitting structures, we make a pixel-by-pixel comparison between the
ALMA image and the MIREX image (Figure \ref{fig:compareflux}).  We
show two different types of comparison. In panel (a), we compare
signal-to-noise ratio (SNR) and $\Sigma$. In panel (b), we compare the
1.3~mm continuum flux density $F_{\rm 1.3mm}$ with $\Sigma$. The
continuum image is primary-beam corrected, so the map boundary regions
have higher noise levels.  Both comparisons are restricted to regions
where the ALMA primary-beam response $\geq$0.5. In both panels, we show the
3$\sigma_{\rm center}$ noise level with a blue dashed horizontal line. The zero
point is shown as the red dashed horizontal line.  
A $3\sigma_{\rm center}$ noise
corresponds to a continuum-derived mass surface density $\Sigma_{\rm
  mm}$ = 0.044 g cm$^{-2}$ \citep[using equation 1 in][]{2017arXiv170105953K}, 
assuming a dust temperature of 20~K,
$\kappa_\nu = 5.95\times 10^{-3}\:{\rm cm^2\:g^{-1}}$ \citep[the
  moderately coagulated thin ice mantle model
  of][]{1994A&A...291..943O}, i.e., with a dust-to-gas mass ratio of
1:141 \citep{2011piim.book.....D}. For a mean particle mass of
2.33$m_{\rm H}$ (i.e., $n_{\rm He}=0.1n_{\rm H}$), this corresponds to
a total column density $N_{\rm H} = 1.9\times10^{22}\:$cm$^{-2}$, i.e,
a visual extinction of $A_V=9.4\:$mag (assuming an extinction to
column density relation $A_V=(N_{\rm H}/2.0\times10^{21}\:{\rm
  cm}^{-2})\:$mag). We note that our restriction of analysis to the
region where the primary-beam correction factor is $<2$ means that
uncertainties associated with this correction are minimized to this
level or smaller.

At first glance, the plots show no clear correlation between the
mm continuum flux and MIREX $\Sigma$.  
A similar situation was found by \citet{2004ApJ...611L..45J}
comparing 0.85 mm continuum emission (observed with JCMT) and near infrared 
extinction (derived from 2MASS data).
However, Figure \ref{fig:compareflux}, shows a hint of detection
deficit of mm continuum emission at $\Sigma\lesssim 0.15 \rm g~cm^{-2}$, although there are
still a modest number of relatively high SNR and flux density values in this
regime. However, one important systematic error associated with the
MIREX map is the presence of MIR-bright sources, which lead to an
underestimation of $\Sigma$ at these locations.  We carry out a visual
identification of potential MIR sources in the Spitzer IRAC $8\:{\rm
  \mu m}$ image and mark their locations in Figure
\ref{fig:cont_mirex}. We then remove these pixels from the analysis,
showing the results in Fig.~\ref{fig:compareflux}(c)(d). There are now
significantly fewer low $\Sigma$ (i.e., $\la0.3\:{\rm g\:cm}^{-2}$)
points with high SNR or flux density values.

Focusing on the results in Fig.~\ref{fig:compareflux}(c)(d), we first
note that there are very few pixels with $\Sigma\la 0.1\:{\rm
  g\:cm}^{-2}$, since even the boundary of the mapped region still
corresponds to quite deeply embedded parts of the molecular
cloud. Also there are relatively few points with $\Sigma\ga 0.6\:{\rm
  g\:cm}^{-2}$, which is partly due to the effects of approaching the
saturation limit in the MIREX map (BTK14). Then, we see that the cloud
of points within $-3\sigma_{\rm center} \la F_{\rm 1.3mm} \la 3\sigma_{\rm center}$ shows
the RMS noise in the continuum image.  At $\Sigma \la 0.5\:\rm
g~cm^{-2}$, most of the pixels still aggregate within $\pm3\sigma_{\rm center}$
RMS noise. However, starting from $\Sigma \sim 0.2\:\rm g~cm^{-2}$, we
see increased numbers of high SNR and flux density values. By $\Sigma
\ga 0.65~\rm g~cm^{-2}$, nearly all points are above the 3$\sigma_{\rm center}$
line. In other words, with the increase of $\Sigma$, it is more likely
to detect 1.3 mm continuum flux with ALMA (given the recoverable
angular scales).  When the IRDC has a high enough mass surface density
($\Sigma \ga 0.65\:\rm g~cm^{-2}$), the 1.3~mm continuum emitting
dense structures are always present. If the continuum detections
indicate current/future star-forming cores, this would indicate that
core/star formation is more likely to happen in high-$\Sigma$ regions
of IRDCs.

\subsection{Dense Gas Detection Probability Function}\label{subsec:dp}

\begin{figure*}[htb!]
\epsscale{0.5}
\plotone{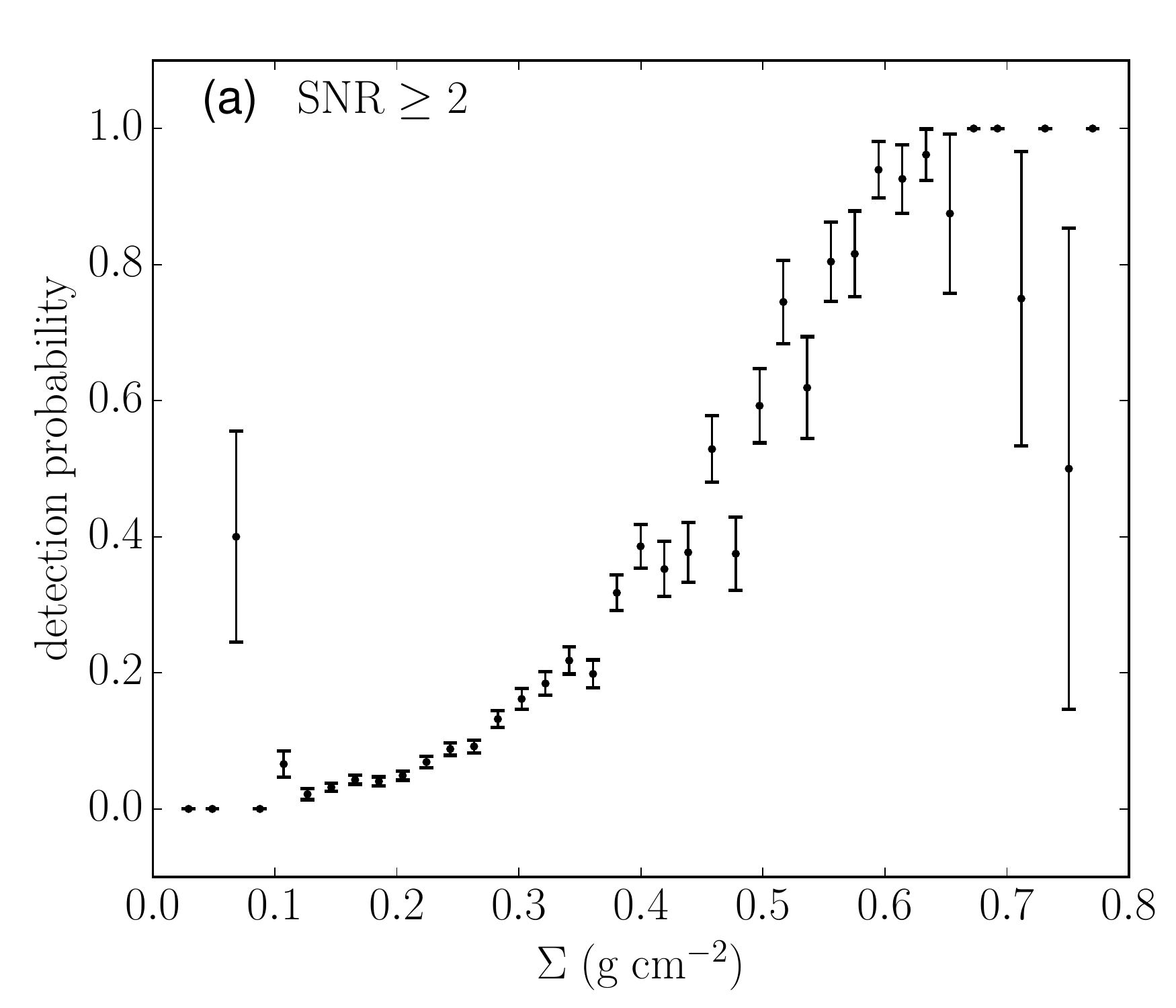}
\plotone{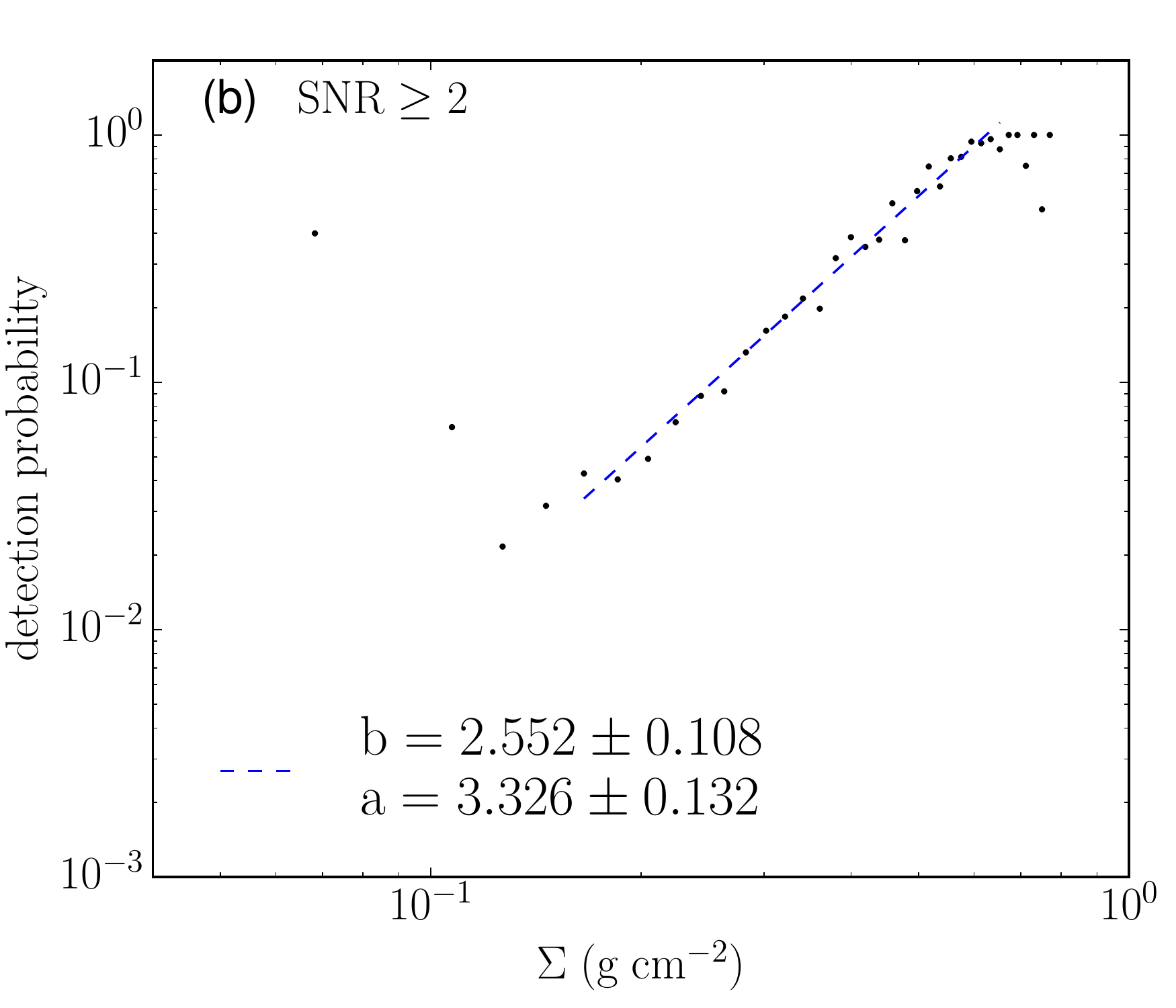}\\
\plotone{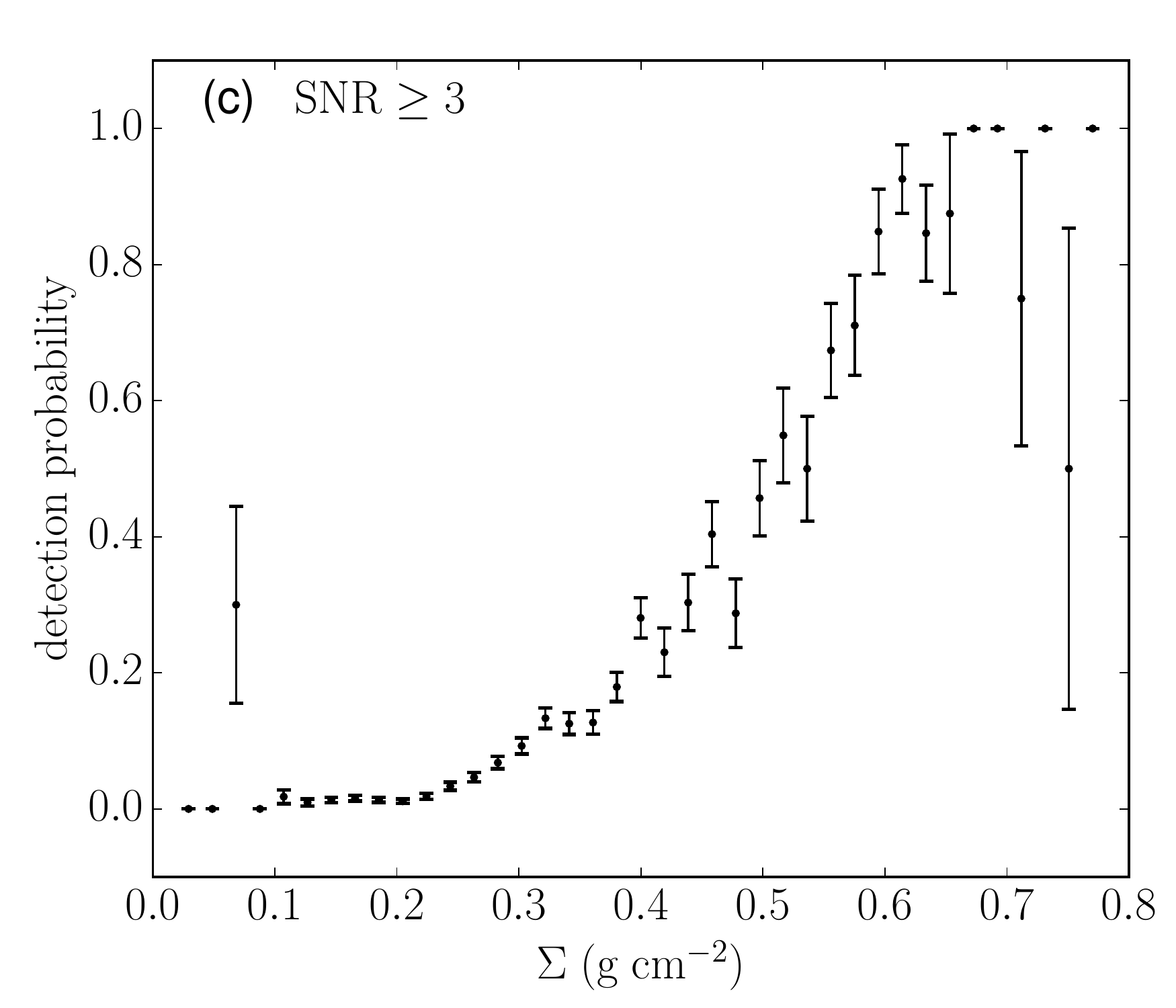}
\plotone{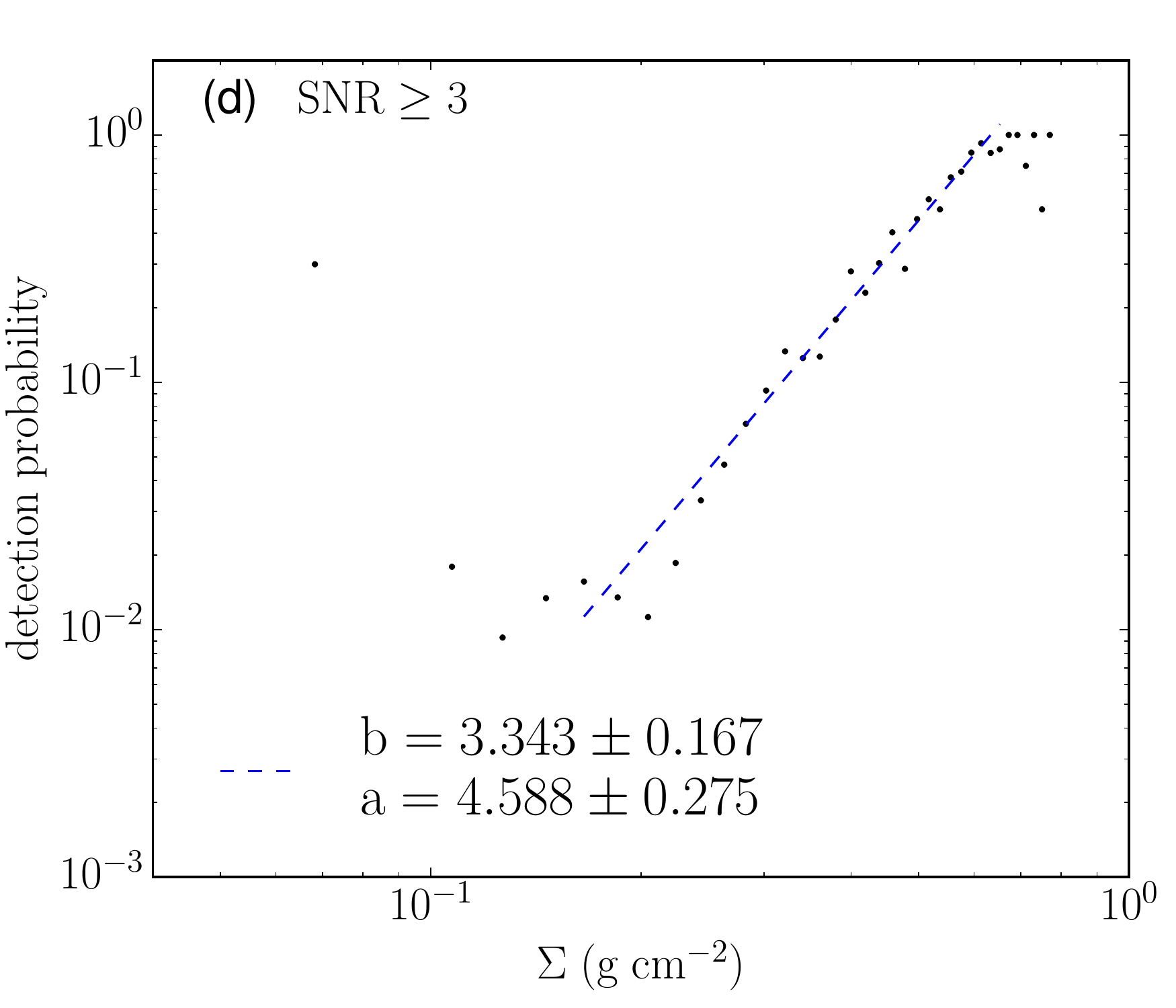}\\
\plotone{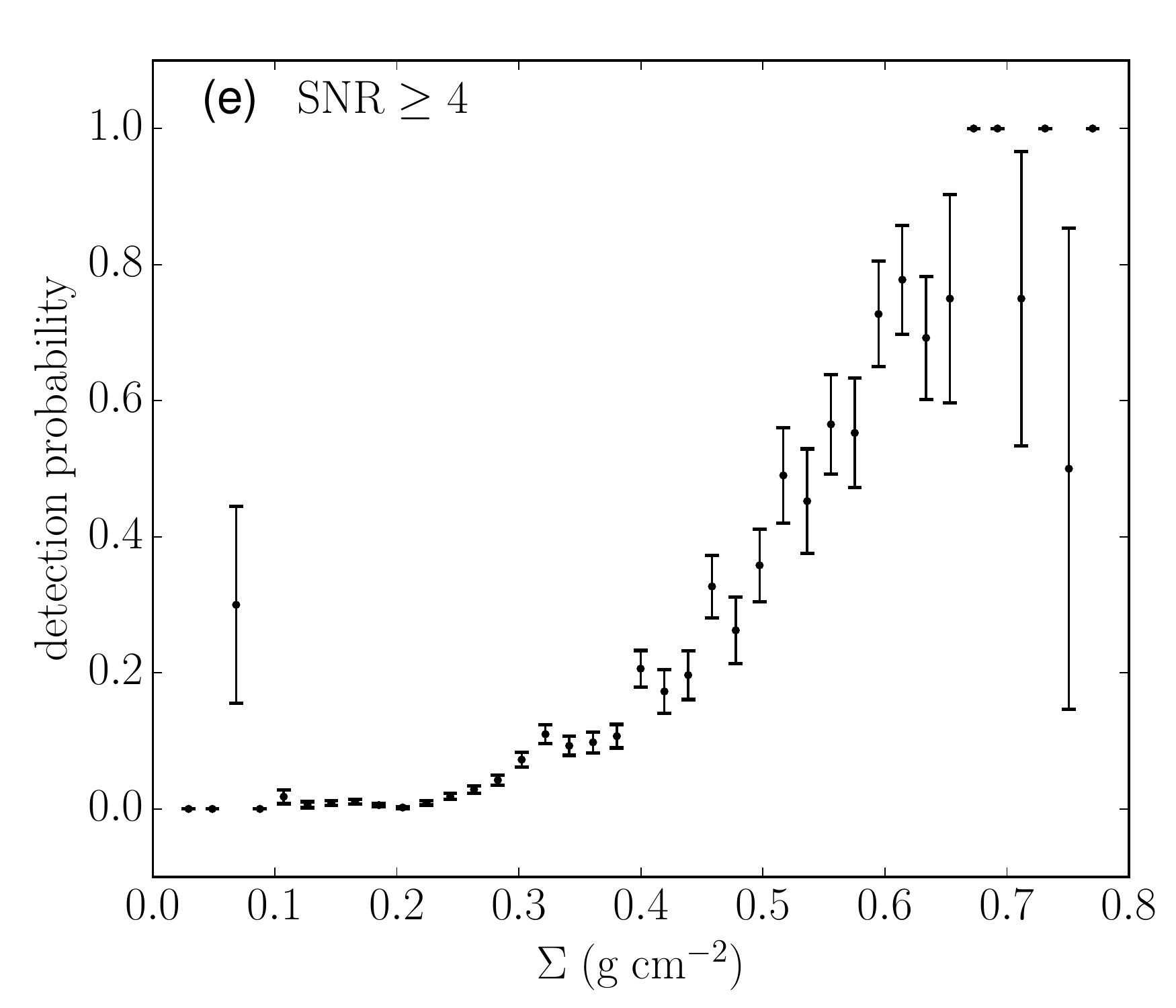}
\plotone{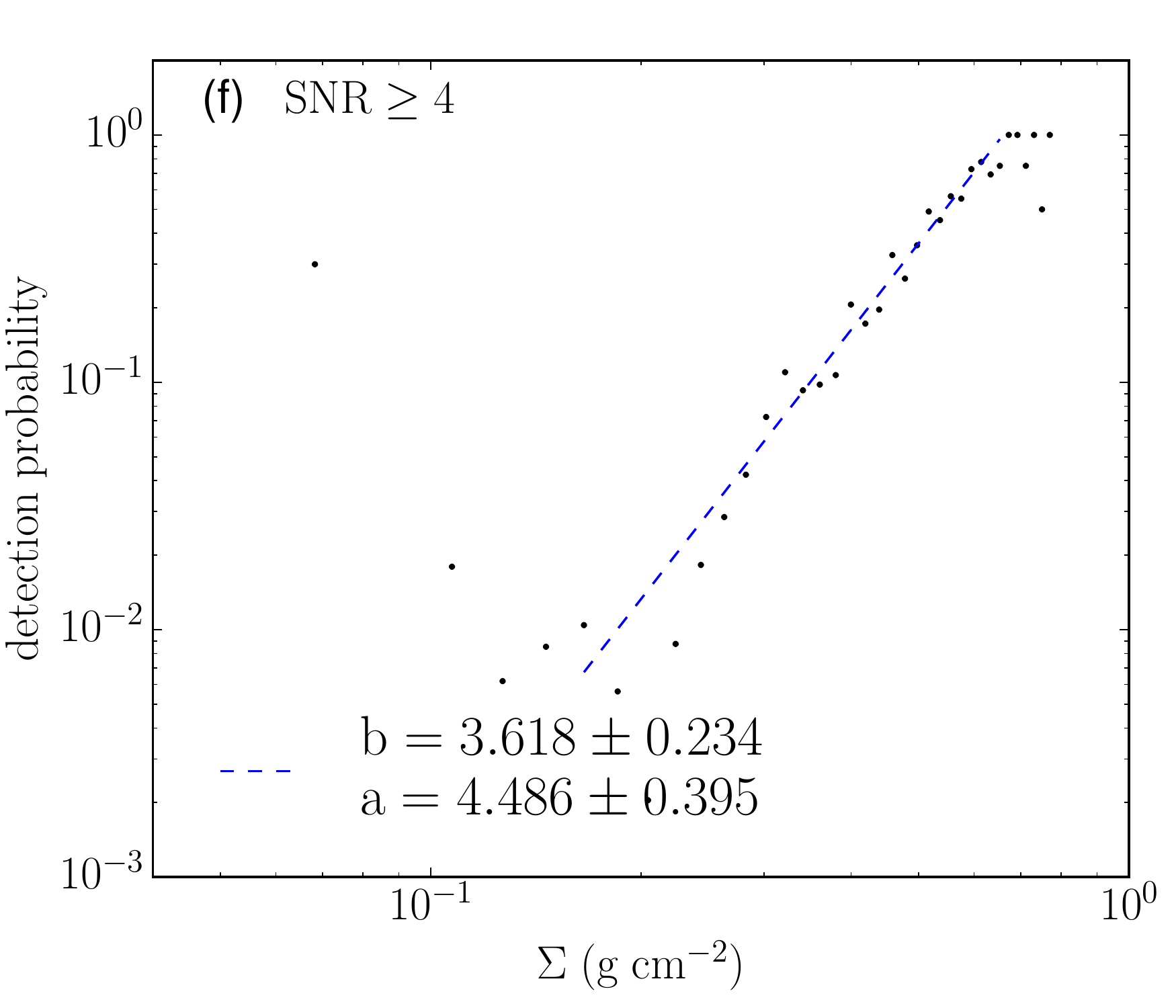}
\caption{
{\bf (a)}: 1.3 mm continuum detection probability 
$P_{\rm 1.3mm}$ as a function of $\Sigma$.
Here the continuum detection threshold is SNR = 2 with 
the RMS noise being 0.2 mJy per 2\arcsec~beam.
The error bars are the square root of the variance of the Bernoulli
distribution (see text).  {\bf (b)}: The same as panel (a), but in
logarithmic scale.  The blue dashed line shows a power-law fit over
the indicated range of $\Sigma$. The parameters $a$ and $b$ follow
equation \ref{eq:dp}. {\bf (c)}: Same as (a), but with a
detection threshold of SNR = 3. {\bf (d)}: Same as (c), but in
logarithmic scale.  {\bf (e)}: Same as (a), but with a
detection threshold of SNR = 4.  {\bf (f)}: Same as (e), but in
logarithmic scale.
\label{fig:dpsnr}}
\end{figure*}

\begin{figure*}[htb!]
\epsscale{0.5}
\plotone{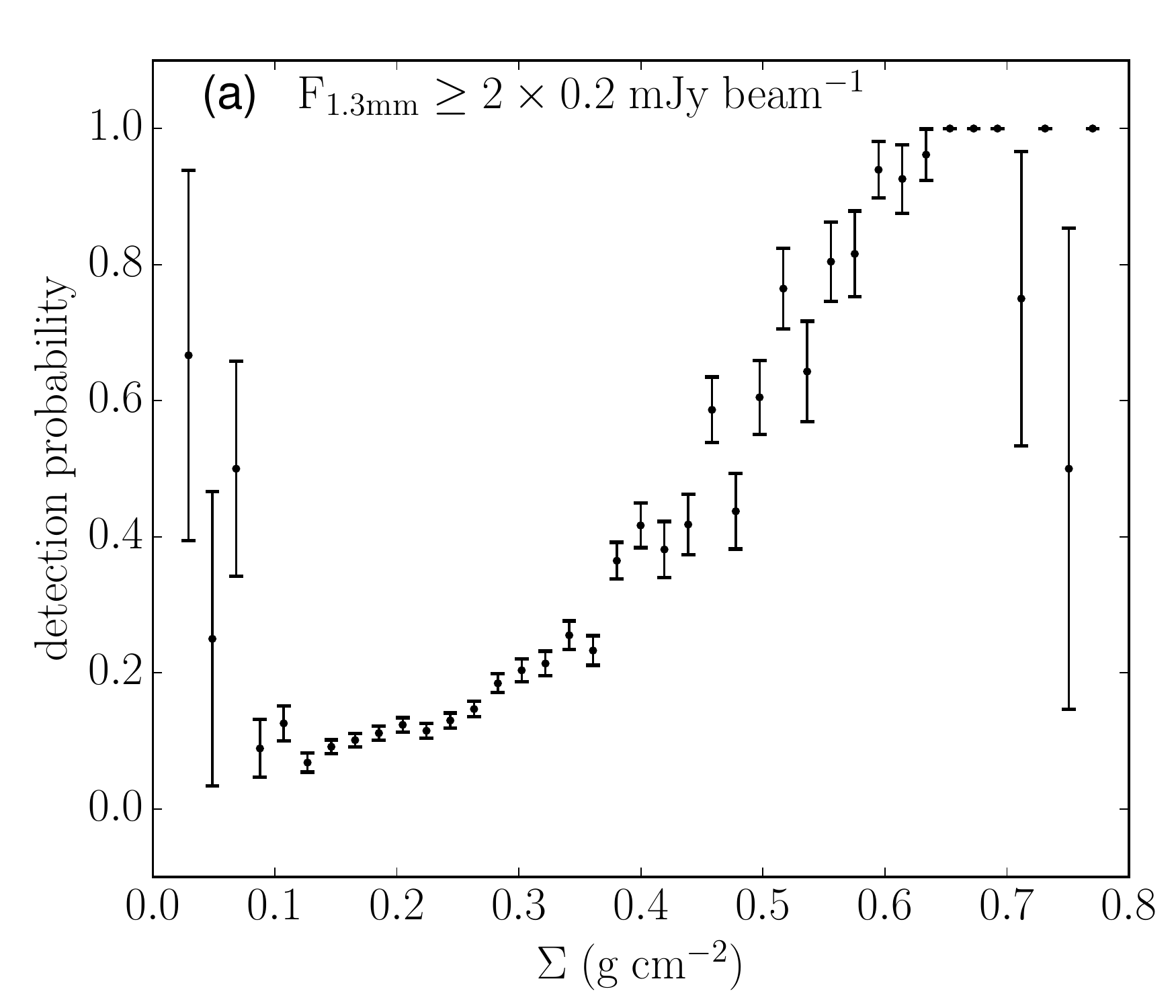}
\plotone{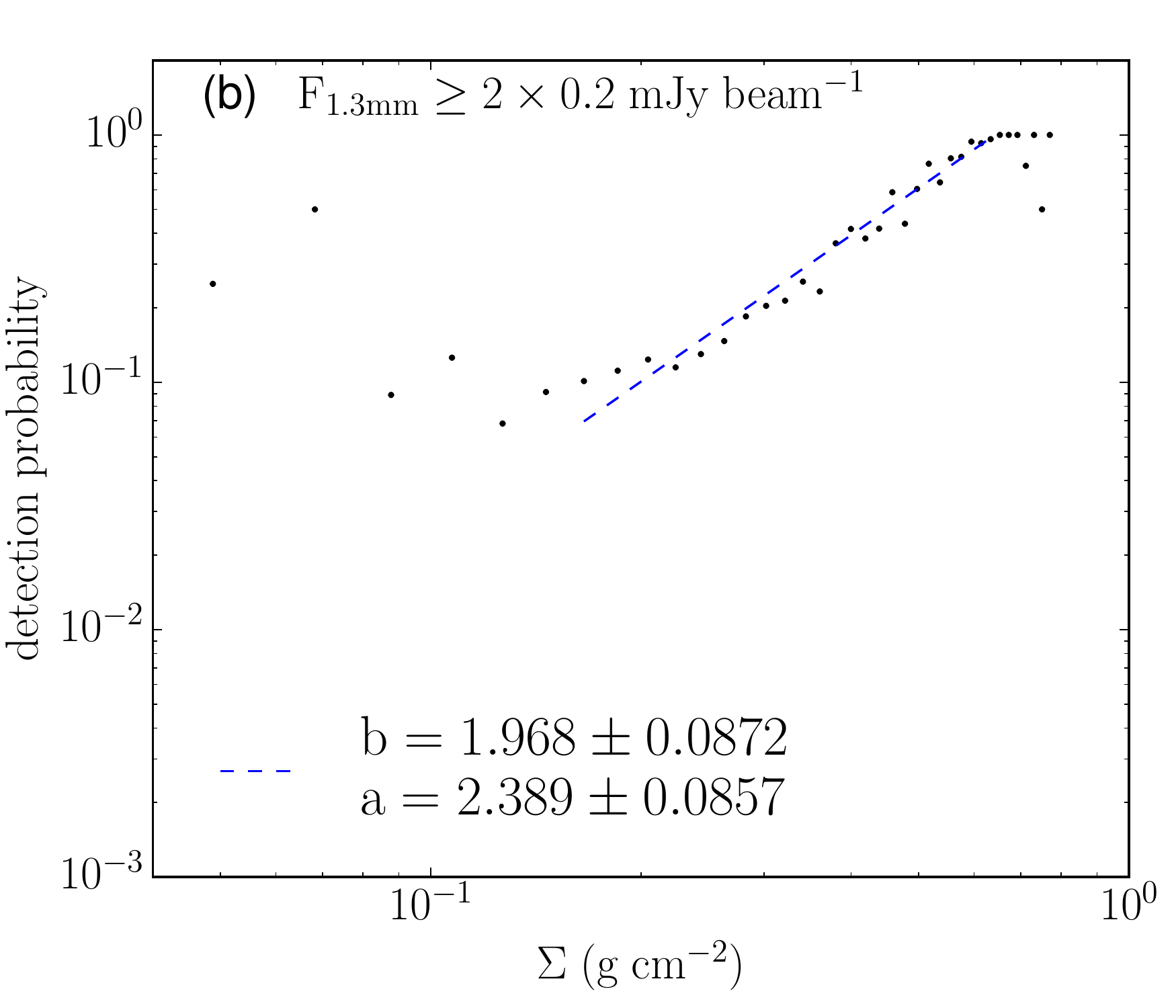}\\
\plotone{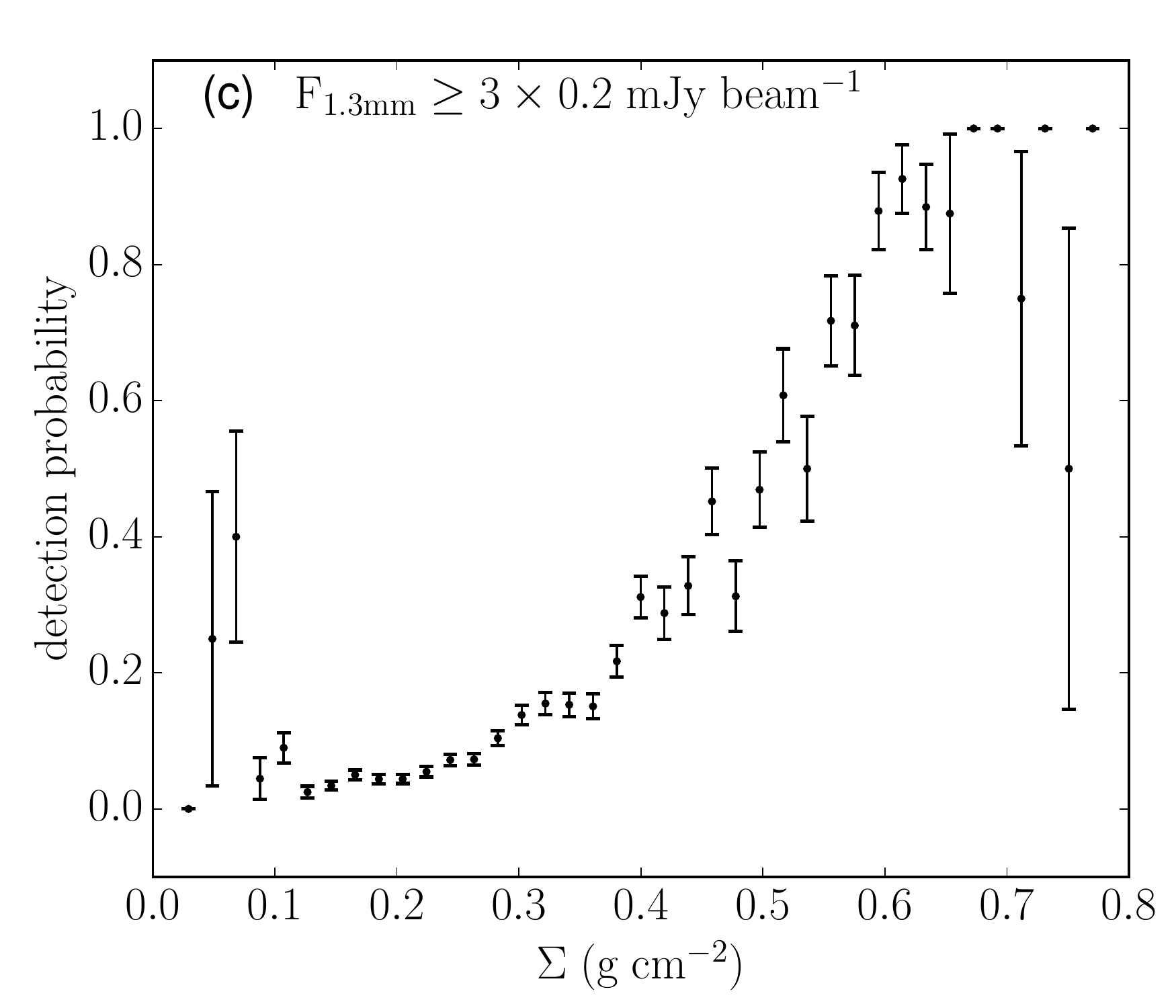}
\plotone{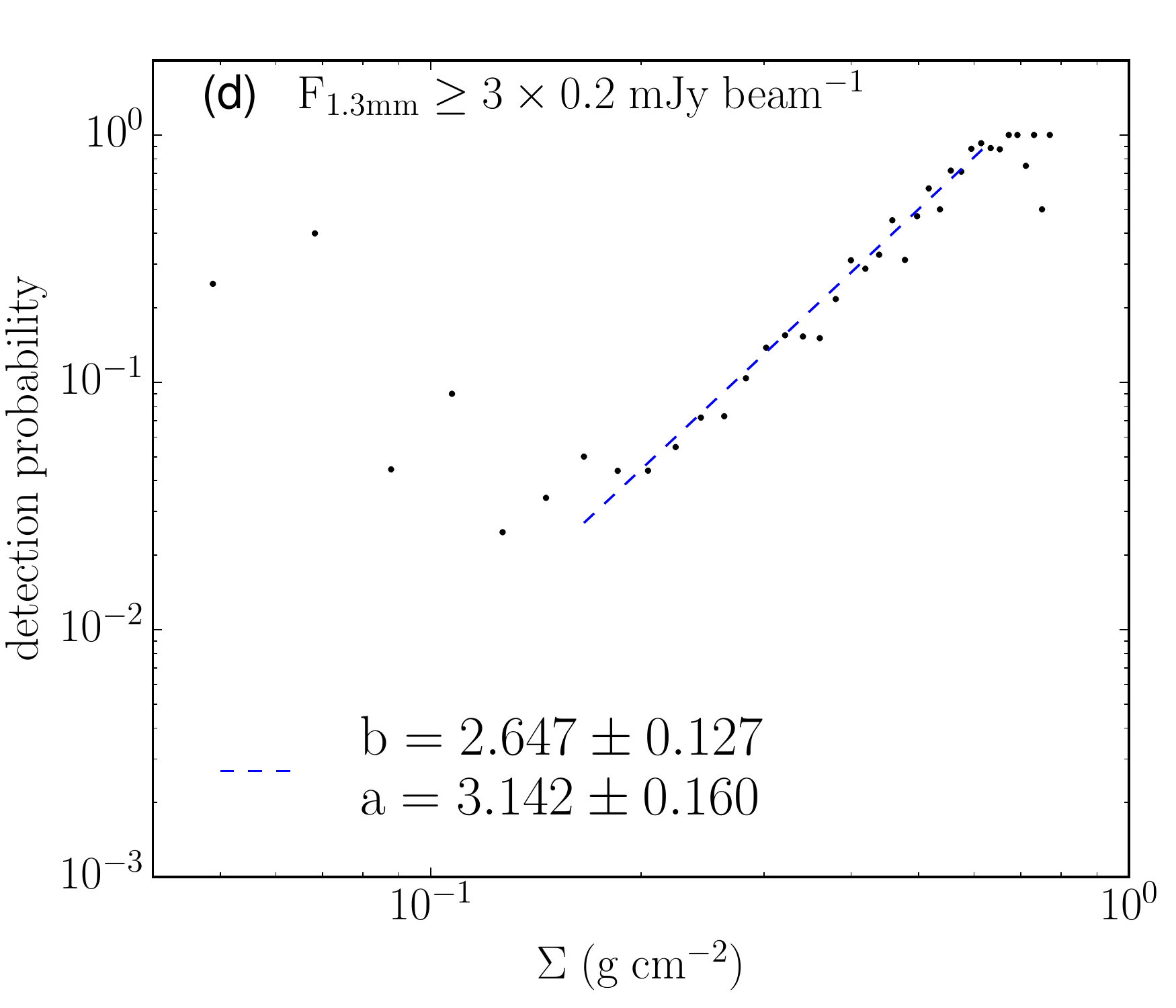}\\
\plotone{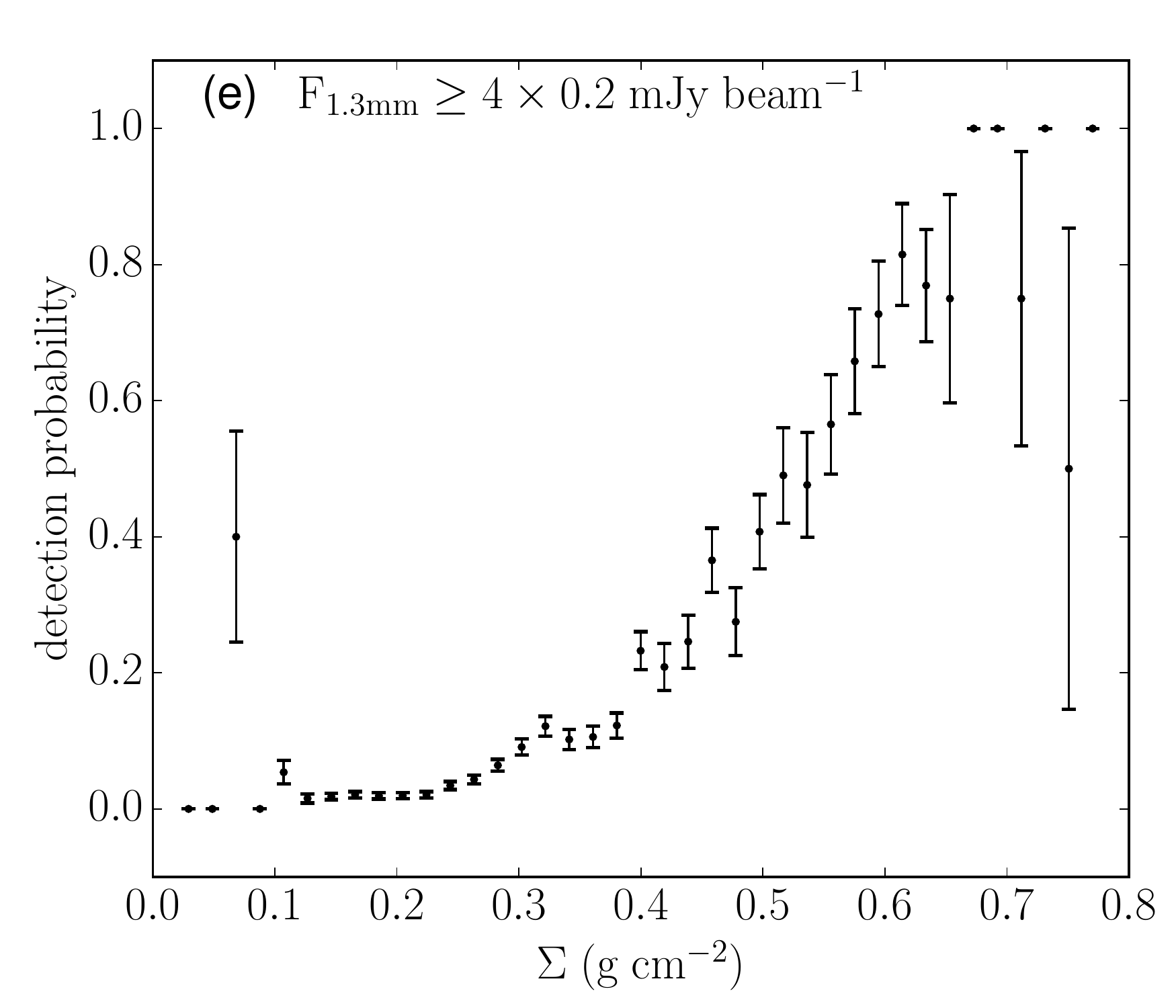}
\plotone{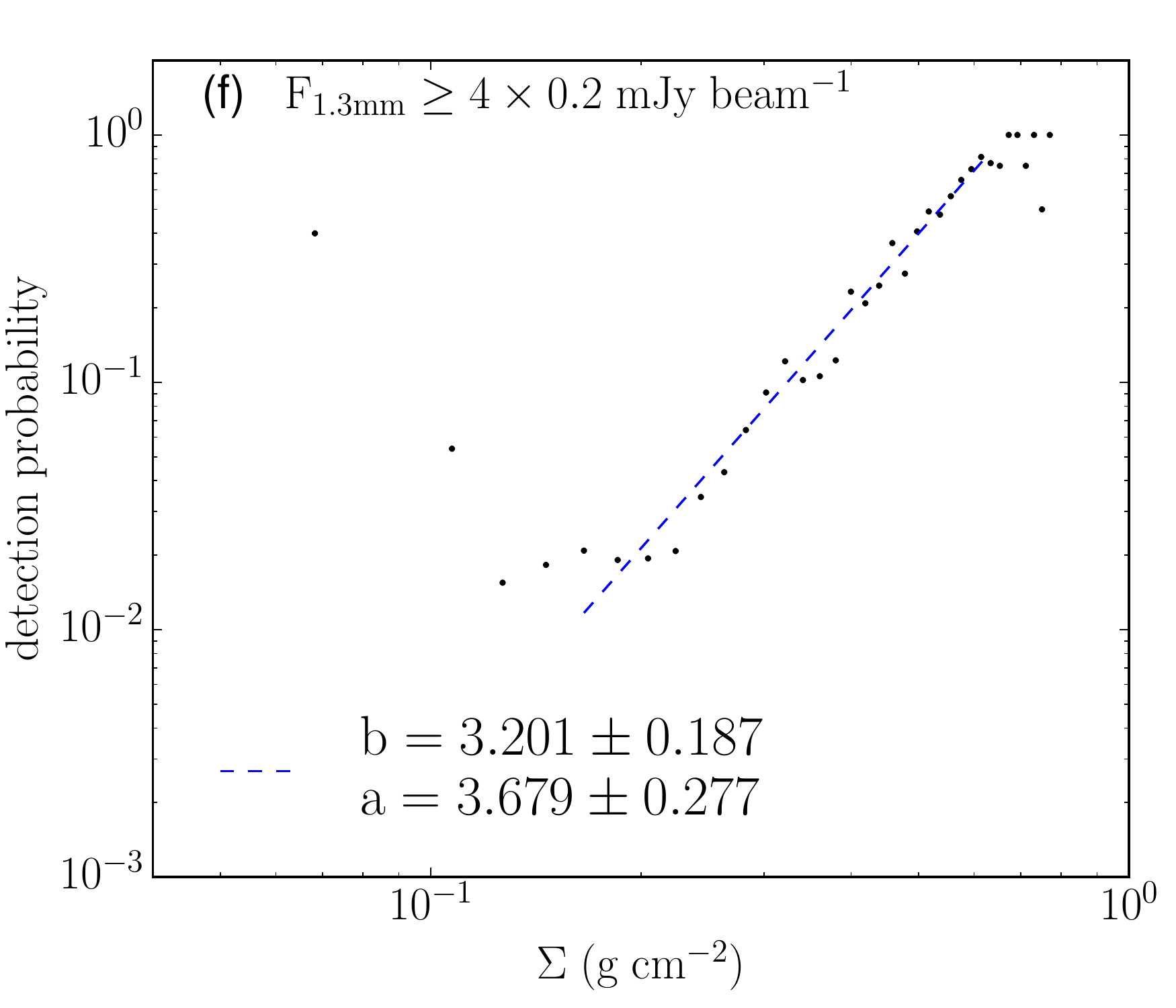}
\caption{
{\bf (a)}: 1.3 mm continuum detection probability
$P_{\rm 1.3mm}$ as a function of $\Sigma$.
Here the continuum detection threshold is 2$\sigma_{\rm center}$,
where $\sigma_{\rm center}$ (= 0.2 mJy per 2\arcsec~beam) is the 
RMS noise at map center (primary-beam response = 1).
The error bars are the square root of the variance of the Bernoulli
distribution (see text).  {\bf (b)}: The same as panel (a), but in
logarithmic scale.  The blue dashed line shows a power-law fit over
the indicated range of $\Sigma$. The parameters $a$ and $b$ follow
equation \ref{eq:dp}. {\bf (c)}: Same as (a), but with a
detection threshold of 3$\sigma_{\rm center}$.
{\bf (d)}: Same as (c), but in logarithmic scale.
{\bf (e)}: Same as panel (a), but with a detection threshold of 4$\sigma_{\rm center}$.
{\bf (f)}: Same as panel (e), but in logarithmic scale.
\label{fig:dpflux}}
\end{figure*}

To further quantify the relation between presence of 1.3~mm continuum
emission and mass surface density of the parent cloud, we plot the
detection probability, $P_{\rm 1.3mm}$, versus $\Sigma$ in Figures
\ref{fig:dpsnr} and \ref{fig:dpflux}, using the dataset with pixels
containing MIR sources removed (see above).  Here $P_{\rm 1.3mm}$ is
defined as the fraction of ``detected'' pixels at a given $\Sigma$.
The definition of detection differs by cases.  In the first case
(Figure \ref{fig:dpsnr}), a pixel is defined to be detected when its
SNR is greater than a given threshold.  A low threshold is more likely
to have false detections, and vice versa.  We adopt a fiducial
threshold of SNR = 3, and show the effects from using SNR = 2 and SNR
= 4.  In the second case (Figure \ref{fig:dpflux}), a pixel is defined
to be detected when its flux density is greater than a given
threshold.  Here we use the primary-beam corrected image.  The
fiducial threshold is 3$\sigma_{\rm center}$ at the map center, where the
primary-beam response is 1.  We also show the effects of using
2$\sigma_{\rm center}$ and 4$\sigma_{\rm center}$.

In the first case of a constant SNR threshold, it is possible that we
miss some weak features at the map boundary where the RMS noise
$\sigma$ is a factor of 2 higher than $\sigma_{\rm center}$.  In the second case
of a constant absolute flux density threshold, while this is closer to
a constant physical limit, i.e., of constant core column density for
fixed dust temperature and dust opacity, the disadvantage is that we
may be overestimating $P_{\rm 1.3mm}$ near the map boundary due to
increased contamination from noise fluctuations.

In these analyses, we adopt a bin size of $\Delta\Sigma$ = 0.02 g
cm$^{-2}$ ($A_V\sim4\:$mag).
In the left columns of Figures \ref{fig:dpsnr} and \ref{fig:dpflux},
we show the $P_{\rm 1.3mm}-\Sigma$ relation with a linear scale.  In
the right columns, we show the relation with a logarithmic scale.
Each row of panels shows the relation with a different detection
threshold, as noted on the top-left corner. 

In each $\Sigma$ bin, $P_{\rm 1.3mm} \equiv N_{\rm detection}/N_{\rm
  total}$.  If each point obeys the Bernoulli distribution with
success probability $p$, i.e.,
\begin{equation}\label{eq:bernoulli}
{\rm P}(X) = 
  \begin{cases}
p & \text{if $X=1$} \\
1-p & \text{if $X=0$} \\
  \end{cases}, 
\end{equation}
where $X=1$ means detection, then $P_{\rm 1.3mm}$ is the expectation
of $\sum_{i=1}^{n}X_i/n$, given $X_1$...$X_n$ are independent,
identically distributed random variables.  The standard deviation of
$\sum_{i=1}^{n}X_i/n$ is $[p(1-p)/n]^{0.5}$, which is adopted as the
error bar for each bin.  We use the observed probability as an
estimate of the Bernoulli success probability. Note that by this
method, estimating the error bar becomes problematic when the success
probability equals 0 or 1. Such points are excluded from the
functional fitting (see below).

At $\Sigma\la 0.04\:\rm g~cm^{-2}$ ($A_V\sim8\:$mag), there are very
few (i.e., only about 5) pixels in the mapped region. While these
pixels do not tend to show mm continuum flux detections via the
various thresholds, there are too few for us to test scenarios of
there being a threshold for star formation at about this level
\citep[e.g.,][]{1989ApJ...345..782M,2004ApJ...611L..45J,2010ApJ...724..687L}.
Also, we note that the MIREX map, even with NIR extinction correction,
can have relatively large systematic errors in this low-$\Sigma$
regime. Indeed, such problems, including incomplete removal of MIR
sources, lead us to be cautious of results for $\Sigma\la 0.15\:\rm
g~cm^{-2}$, where $P_{\rm 1.3mm}$ is seen to sometimes have finite
values, but typically with large errors.

However, in the main region of interest for our study, i.e., for
$\Sigma\ga 0.15\:\rm g~cm^{-2}$, in all the cases the detection
probability increases steadily to reach approximately 100\% by
$\Sigma\sim 0.65~\rm g~cm^{-2}$.  In Figures \ref{fig:dpsnr}(b)(d)(f)
and \ref{fig:dpflux}(b)(d)(f), the plots show that $P_{\rm 1.3mm}$
follows an approximate power-law relation with $\Sigma$ between
$\Sigma\sim 0.15~\rm g~cm^{-2}$ and $\Sigma\sim 0.65~\rm g~cm^{-2}$.
We fit the function $P_{\rm 1.3mm}=a(\Sigma / 1\:\rm g~cm^{-2})^b$ by
minimizing $\chi^2$ (normalized by the errors), which is shown as the
blue dashed lines in these figures.
Note, we do not include $P_{\rm 1.3mm}$ = 1 points in the fit.
The resulting power-law indices $b$ and amplitudes $a$ are displayed in
the figures and in Table~\ref{tab:dpsum}.

With an increase in the level of the detection thresholds, Figures
\ref{fig:dpsnr} and \ref{fig:dpflux} show a decrease in detection
probabilities, as expected.  At the same time, the power-law indices
become larger, i.e., with a higher detection threshold, the increase
of $P_{\rm 1.3mm}$ between $\Sigma\sim 0.15~\rm g~cm^{-2}$ and
$0.65~\rm g~cm^{-2}$ becomes steeper.  
In the next section we will use such power law approximations for
$P_{\rm 1.3mm} (\Sigma)$ to estimate the mass fraction of ``dense''
gas in the IRDC and GMC region.

\begin{table*}
\centering
\begin{threeparttable}
\large
\caption{Detection Probability Relations}\label{tab:dpsum}
\begin{tabular}{cccccccc}
\hline {{thresholds}} & {{$\Sigma_{\rm mm}\rm (g~cm^{-2})$}} & {{$a$}}& {{$b$}} & {{$f_{\rm dg,mm}$}} & {{$f_{\rm dg,MIREX}$}} & {{$f_{\rm dg,DPF,0.15-0.65}$}} & {{$f_{\rm dg,DPF,GMC}$}} \\
{{(1)}} & {{(2)}} & {{(3)}} & {{(4)}} & {{(5)$^{(a)}$}} & {{(6)}} & {{(7)}} & {{(8)}}\\
\hline
SNR$\geq$2 & 0.029 & 3.3 & 2.6 & 10\%$_{6.0\%}^{15\%}$  & 20\% & 17\% &  9.2\% \\  
SNR$\geq$3 & 0.044 & 4.6 & 3.3 & 8.7\%$_{5.3\%}^{13\%}$ & 13\% & 12\% &  6.5\% \\  
SNR$\geq$4 & 0.058 & 4.5 & 3.6 & 8.0\%$_{4.9\%}^{12\%}$ & 10\% & 9.4\% & 5.7\% \\  
\hline
\hline
$F_{\rm 1.3mm}\geq2\sigma_{\rm center}$ & 0.029 & 2.4 & 2.0 & 11\%$_{6.5\%}^{16\%}$  & 24\% & 22\% & 13\% \\  
$F_{\rm 1.3mm}\geq3\sigma_{\rm center}$ & 0.044 & 3.1 & 2.6 & 9.5\%$_{5.8\%}^{14\%}$ & 17\% & 15\% & 8.2\% \\  
$F_{\rm 1.3mm}\geq4\sigma_{\rm center}$ & 0.058 & 3.7 & 3.2 & 8.6\%$_{5.2\%}^{13\%}$ & 12\% & 11\% & 6.3\% \\  
\hline
\hline
\end{tabular}
\begin{tablenotes}
\small
\item (a) The super- and subscripts correspond to 
using the lower (15 K) and higher (30 K) temperature assumptions
in the mass estimation based on 1.3 mm continuum flux.
\end{tablenotes}
\end{threeparttable}
\end{table*}

\subsection{Dense Gas Fraction}\label{subsec:efficiency}

The ALMA observations give us a direct measure of the amount of
``dense'' gas, i.e., that is detected by some defined criteria of
1.3~mm flux emission, which can be compared to the total mass estimate
of the IRDC that overlaps with the region mapped by ALMA. From the
MIREX map, this mass is $1.21\times10^4\:M_\odot$, with
uncertainties at the level of about 30\% due to opacity per unit total
mass uncertainties. Distance uncertainties contribute further, but
these will cancel out in the ratio of these masses to the mm-continuum
derived mass.

The total mm flux in the observed, analyzed region (i.e.,
where primary beam correction factor is $\leq 2$) is 1.42~Jy (based on
detections above $3\sigma_{\rm center}$), which translates into a
total mass of $1.16\times10^3\:M_\odot$ given our fiducial
assumptions, including $T=20\:$K. Thus the direct measure of dense gas
mass fraction (expressed as percentages) is $f_{\rm dg,mm}= 9.5\%$ for
this case.  This value is listed in column (5) of
Table~\ref{tab:dpsum} for all the considered cases, and showing the
effects of varying $T$ from 15~K to 30~K. We see the sensitivity of
these dense gas fractions to threshold choice and temperature choice,
with fiducial results being about 10\%.
Systematic variations arising
from the choice of dust temperature are up to a factor of almost two
and are the most significant source of uncertainty \citep[see also][]{2009ApJ...692...91G}.

A second estimate of the dense gas fraction, $f_{\rm dg,MIREX}$ can be
made by summing the MIREX mass estimate of the pixels that are
detected in 1.3~mm continuum. These values are shown in column (6) of
Table~\ref{tab:dpsum}. Fiducial results are now moderately higher at
about 15\%. 

Next we utilize our analytic approximations for the detection
probability function (DPF), $P_{\rm 1.3mm}(\Sigma)$, combined with
analytic forms for the probability distribution function (PDF) of
$\Sigma$ to estimate dense gas fractions. Recall, the observed DPFs
have a power law form in the range from $\Sigma\sim 0.15 \rm
g~cm^{-2}$ to $\sim0.65 \rm g~cm^{-2}$. At lower values of $\Sigma$ we
extrapolate with a constant that is 
similar to the $P_{\rm 1.3mm}$ at $\Sigma= 0.15\:
\rm g~cm^{-2}$. Finally at high values, $\Sigma>0.65 \rm g~cm^{-2}$ we
use a constant value of unity. Thus, overall the DPF is described via
\begin{equation}\label{eq:dp}
P_{\rm 1.3mm} = 
  \begin{cases}
P_{\rm 1.3mm,min} & \text{if $\Sigma/({\rm g~cm}^{-2})< 0.15$} \\
a(\frac{\Sigma}{1~\rm g~cm^{-2}})^{b} & \text{if $0.15<\Sigma/({\rm g~cm}^{-2}) <0.65$} \\
1 & \text{if $\Sigma/({\rm g~cm}^{-2}) > 0.65$}, \\
  \end{cases} 
\end{equation}
where $P_{\rm 1.3mm,min} = a(0.15~\rm g~cm^{-2} / 1~\rm
g~cm^{-2})^{b}$. The fiducial value for $P_{\rm 1.3mm,min}$ is
$\sim0.02$. This value acts effectively as a lower limit floor on our
estimated values of $f_{\rm dg}$.

Then the mass of dense, i.e., 1.3mm-emitting, gas is 
\begin{equation}\label{eq:coremass}
M_{\rm dg} = \int P_{\rm 1.3mm} \Sigma A p({\rm ln}\Sigma) {\rm d(ln}\Sigma)
\end{equation}
where $A$ is the total cloud area being integrated over and $p({\rm
  ln}\Sigma)$ is the cloud's PDF of mass surface densities.

Based on two independent methods, the $\Sigma$-PDF in IRDC G28.37+0.07
and its surroundings (i.e., of a $\sim20\arcmin$-scale region,
equivalent to $\sim30\:$pc) has been found to be reasonably well fit
by a single log-normal function
\citep{2014ApJ...782L..30B,2016ApJ...829L..19L}, i.e., of the form
\begin{equation}\label{eq:SigmaPDF}
p(\rm ln\Sigma) = \frac{1}{\sqrt{2\pi}\sigma_{\rm ln\Sigma}}\rm exp[-\frac{(\rm ln\Sigma-\overline{\rm ln\Sigma})^2}{2\sigma_{\rm ln\Sigma}^2}].
\end{equation}
Here we adopt this empirical $\Sigma$-PDF (i.e., area-weighted)\footnote{We 
have also made the same calculations using their mass-weighted PDF. The
results (dense gas fractions) are very similar.} in the
NIR+MIR extinction map case, i.e., with $\sigma_{\rm ln\Sigma}$ = 1.15,
$\overline{\rm \Sigma}$ = 0.038 $\rm g~cm^{-2}$, and $\overline{\rm
  ln\Sigma}$ = -3.93. We note that the actual $\Sigma$-PDF measured by
\citet{2016ApJ...829L..19L} has a small power law tail excess
component, emerging at about $\Sigma\sim0.3\:{\rm g\:cm}^{-2}$. While
the use of the above log-normal leads to a small underestimation of
the importance of the higher $\Sigma$ regions, it is a very modest
effect since the fraction of pixels affected by this excess is
less than a few percent.

Then the total mass of dense gas can be estimated by integrating
equation \ref{eq:coremass}. If we carry out this exercise for the area
corresponding to the analyzed area of the IRDC, i.e., that mapped by
ALMA with a primary beam response $>0.5$, we obtain
135$\:M_\odot$. This is much smaller than our previous estimates for
$M_{\rm dg}$, which is primarily because the $\Sigma$-PDF was
estimated for a much larger region and contains much more contribution
from lower values of $\Sigma$. If we restrict the above integration to
the range $\Sigma=0.15$ to $0.65\:{\rm g\:cm}^{-2}$, then we obtain
$M_{\rm dg}=1,828\:M_\odot$ (for the $>3\sigma_{\rm center}$ case), in
much closer agreement with our previous estimates.  Dense gas
fractions calculated via this latter method can be derived by
comparison to the total cloud mass observed in the mapped region,
i.e., $1.21\times10^4\:M_\odot$, yielding the values $f_{\rm
  dg,DPF,0.15-0.65}$ in column (7) of Table~\ref{tab:dpsum}.  These
values are very similar to those of $f_{\rm dg,MIREX}$.

Finally, we can make the extrapolation that the observed DPF of the
inner IRDC region mapped by ALMA will hold in the wider GMC region,
where the approximately log-normal $\Sigma$-PDF was measured. For this
$\sim30\:$pc-scale region, the total cloud mass is
\begin{equation}\label{eq:cloudmass}
M_{\rm tot} = \int \Sigma A p({\rm ln}\Sigma) {\rm d(ln}\Sigma),
\end{equation}
which has a value of 170,000$\:M_\odot$. The values of $f_{\rm
  dg,DPF,GMC}=M_{\rm dg}/M_{\rm tot}$ are shown in column (8) of
Table~\ref{tab:dpsum}. In the fiducial cases, these values are
smaller than 10\%.

\section{Discussion}\label{sec:discussion}

\subsection{Core/Star Formation Efficiency}

The MIR extinction map and the ALMA 1.3~mm continuum map both trace
dust in the IRDC, which are then used to estimate the masses. However,
while the MIREX map traces the total mass surface density without bias
at any particular spatial scale and without bias on the temperature
(as long as the region is cold enough not to be emitting at $\rm
8\:\mu m$), the ALMA continuum map misses flux from extended
structures ($\gtrsim20\arcsec$) and is biased towards warmer
material. We describe the mass associated with the 1.3~mm continuum
flux as the ``dense'' gas component and discuss below that the
majority of this material is likely to be directly involved in the
star formation process.

We have measured the mass of the component that is detected by our
ALMA observation of dust continuum emission and find it to be about
$f_{\rm dg,mm}\sim$10\% of the total mass in the ``central'', i.e.,
mapped region of the IRDC, but with about 50\% uncertainties due to
assumed dust temperature.  If we use the values of the MIREX pixels at
the locations where mm continuum emission is seen, then the associated
mass fraction increases by a factor of about 1.7 (depending on the
choice of flux threshold), i.e., to $f_{\rm dg,MIREX}\sim$17\%. The
difference between $f_{\rm dg,MIREX}$ and $f_{\rm dg,mm}$ could be due
to, e.g., a dense core filling factor of less than one on the scale of
the 2\arcsec\ pixels or a systematically lower temperature of the mm
continuum emitting dust, i.e., $\sim15\:$K rather than $20\:$K.

If we use the data to define a detection probability of mm continuum
emission as a function of $\Sigma$ and then apply this to an estimate
of the $\Sigma$-PDF of the mapped region of the IRDC, i.e., a
log-normal but restricted to the range of $\Sigma=0.15$ to $0.65\:{\rm
  g\:cm}^{-2}$, then we obtain values of dense gas fractions of
$f_{\rm dg,DPF,0.15-0.65}\simeq$15\%, very similar to the values of
$f_{\rm dg,MIREX}$ (also compare other values in columns 6 and 7), 
which indicates that the analytic approximations
for the DPF are quite accurate. Extrapolating the observed DPF of the
inner IRDC region to the wider GMC region, where the $\Sigma$-PDF was
seen to be well-fit by a single log-normal
\citep[BTK14;][]{2016ApJ...829L..19L}, then integration with this PDF
leads to estimates of $f_{\rm dg,DPF,GMC}\simeq$8\%. We note that this
mass fraction is very similar to the mass fraction of the GMC that is
in the power law tail part of the $\Sigma$-PDF, $\epsilon_{\rm
  pl}\sim$3\% to 8\% \citep[][]{2016ApJ...829L..19L} based on lower
angular resolution Herschel measurements of sub-mm dust continuum
emission from the region.

Our ALMA continuum map detects the C1a, C1-Sa, and C1-Sb protostellar
cores from \citet{2016ApJ...821L...3T}, which includes some lower-mass
objects. It also detects the five main continuum structures in C2
\citep{2009ApJ...696..268Z,2015ApJ...804..141Z}, which have been
resolved into a population of cores extending down to sub-solar
masses. Thus it is likely that the current observations capture a
significant fraction of the core mass function (CMF) of protostellar
cores. The detected mm flux may also contain some contribution from
more massive pre-stellar cores, such as C1-S and C1-N
\citep{2013ApJ...779...96T,2017arXiv170105953K}. Thus, for
simplicity, we will assume that our detected 1.3~mm continuum fluxes
give a near complete census of the protostellar CMF and ignore the
possibility that it may include some contribution from the pre-stellar
CMF. These effects of protostellar CMF incompleteness and pre-stellar
CMF contribution will offset each other to some extent. Under this
assumption, then the total current star-forming core efficiency is
simply the same as $f_{\rm dg}$. If we next further assume that the
star formation efficiency from individual cores is about 50\%, which
is expected based on models of outflow feedback
\citep{2000ApJ...545..364M,2014ApJ...788..166Z}, then the total mass
of stars that would form from the currently observed cores is about
half of $f_{\rm dg}$, i.e., $\epsilon_*\sim$5\% to $\sim$8\%.

\subsection{Star Formation Rates}

A number of star formation models involve protostellar cores
collapsing at rates similar to their local free-fall rate
\citep[e.g.,][]{1987ARA&A..25...23S,2003ApJ...585..850M,2005ApJ...630..250K}.
The Turbulent Core Model \citep[][hereafter MT03]{2003ApJ...585..850M} 
assumes core properties are set by
the mean pressure in their surrounding, self-gravitating clump, which
then leads to a simple relation between the individual star formation
time and the average free-fall time of the clump.  In the fiducial
case the timescale for star formation is $t_{*f} = 1.3\times 10^{5}
(M_{c}/60\:M_\odot)^{1/4} (\Sigma_{\rm cl}/1\:{\rm
  g\:cm}^{-2})^{-3/4}\:{\rm yr}$ (cf. equation 44 in MT03), 
which has a very weak dependence on
core mass, $M_c$, and clump mass surface density, $\Sigma_{\rm
  cl}$. This timescale is related to the clump's mean free-fall time
via $t_{*f}/\bar{t}_{\rm ff,cl}=0.98(M_c/60\:M_\odot)^{1/4}(M_{\rm
  cl}/4000\:M_\odot)^{-1/4}$ (cf. equation 37 in MT03), 
i.e., they are quite similar.

For a CMF that is a Salpeter (1955) power law of form $dN/d{\rm
  log}M_c\propto M_c^{-\alpha}$ with $\alpha=1.35$ with lower limit of
$M_c=1\:M_\odot$ and upper limit of $240\:M_\odot$ (so that resulting
stellar IMF with 50\% formation efficiency from the core is in the
range from $m_*=0.5\:M_\odot$ to $120\:M_\odot$, which is, for our
purposes, a reasonable approximation of the actual observed IMF), then
half of the mass of the core population has
$M_c\gtrsim5\:M_\odot$. Thus we take $M_c=5\:M_\odot$ as a typical
core mass. The mapped region of the IRDC has a total mass of
$\simeq1.21\times10^4\:M_\odot$, which we will approximate as
$10^4\:M_\odot$. Under these two conditions, $t_{*f}/\bar{t}_{\rm
  ff,cl}\rightarrow0.42$.

Assuming the SFR is steady and the CMF is evenly populated, then the
observed cores will represent those objects that have formed in the
last average individual star formation time, $\bar{t}_{*f}$, i.e., the
last $0.42\bar{t}_{\rm ff,cl}$. Taking the mass fraction in dense gas
(defined at $F_{\rm 1.3mm}\geq3\sigma_{\rm center}$) as $f_{\rm
  dg,mm}=0.095^{0.14}_{0.058}$ as the most accurate estimate of the
current mass fraction in protostellar cores in the observed region of
the IRDC, then we find that, for $\epsilon_{\rm core}=0.5$ 
\citep{2000ApJ...545..364M,2014ApJ...788..166Z}, the star formation
efficiency per free-fall time is $\epsilon_{\rm
  ff}=0.11^{0.17}_{0.069}$.

This estimate of $\epsilon_{\rm ff}\sim 0.1$ is about a factor of two
larger than the value estimate inside the half-mass radius of the
Orion Nebula Cluster by \citet{2014ApJ...795...55D}, which was estimated from
observed age spreads of young stellar objects. However, the
uncertainties arising solely from the uncertain temperatures of
protostellar cores (15 to 30~K range adopted here) lead to almost
a factor of two uncertainty in $\epsilon_{\rm ff}$. The mean mass
surface density in the analyzed region of the IRDC is
$\simeq0.26\:{\rm g\:cm}^{-2}$. The protostellar core models of 
\citet{2015ApJ...802L..15Z}, i.e., for $M_c=2, 3, 4\:M_\odot$, in
$\Sigma\simeq0.3\:{\rm g\:cm}^{-2}$ clump environments have mean
envelope temperatures near 20~K (set mostly by accretion
luminosities), but can exceed 30~K in $\Sigma\simeq1\:{\rm
  g\:cm}^{-2}$ regions that have higher accretion rates. Also, more
massive cores forming more massive protostars, will tend to have
warmer envelope temperatures, which would lower our estimates of the
mass of the core population and thus $\epsilon_{\rm ff}$. These
uncertainties can be reduced by carrying out temperature measurements
of each protostellar core (e.g., of the dust via spectral energy
distribution observations and modeling or of associated gas via, e.g.,
$\rm NH_3$ observations).

In addition to the effects of core temperature uncertainties,
additional systematic uncertainties include that the analysis has
assumed a fixed value of the star formation efficiency from the core,
a particular relation between star formation time and clump free-fall
time (fiducial case from MT03) and equates the observed 1.3~mm
continuum structures with the total protostellar core
population. These assumptions and uncertainties can be improved with
future work. For example, observations of CO outflows can be used to
confirm that mm continuum sources are indeed protostellar
cores. Better sensitivity of mm continuum data can help to probe
further down the protostellar CMF (although with the half-mass point
estimated to be near $5\:M_\odot$, we expect that the bulk of the
population containing most of the mass has already been
detected). Assumptions about star formation efficiency from the core
can be tested with improved theoretical and numerical models
\citep[e.g.,][]{2017ApJ...835...32T,2017MNRAS.470.1026M}. The relation of
individual star formation time to mean clump free-fall time is more
difficult to test observationally, and may depend on the uncertain
degree of magnetization in the cores \citep[][MT03]{1997ApJ...475..237L}. One
observational test involves measuring the mass accretion rates of the
protostars, potentially from modeling their spectral energy
distributions \citep[see, e.g.,][]{2017ApJ...843...33D,2015ApJ...802L..15Z,2017arXiv170808853Z}
or from measuring their mass outflow rates that are expected to
be proportional to accretion rates \citep[see, e.g.,][]{2016A&ARv..24....6B}.

\section{Conclusions}

In this paper, we have presented first results from an ALMA 1.3~mm
continuum mosaic observation using the 12-m array of the central
regions of a massive IRDC, which is a potential site of massive star
cluster formation. We have focused on carrying out a detailed
comparison of the 1.3~mm emission (which is sensitive to structures
$\la20\arcsec$ in size) with a MIR-derived extinction map of the
cloud. In particular, we argue that the 1.3~mm structures likely trace
``dense'', protostellar cores, and have studied the prevalence of such
sources in the IRDC as a function of its local mass surface density,
$\Sigma$. Based on various definitions of 1.3~mm continuum detection,
i.e., at a fixed signal to noise ratio or a fixed absolute flux
density, we find that the detection probability function (DPF),
$P_{\rm 1.3mm}(\Sigma)$, rises as a power law, i.e., $\propto
(\Sigma/1\:{\rm g\:cm}^{-2})^b$ with $b\sim3$ in the fiducial cases,
over the range $0.15 \la \Sigma / 1\:{\rm g\:cm}^{-2} \la 0.65$. At
higher values of $\Sigma$, we find that $P_{\rm 1.3mm}\simeq 1$. At
lower values of $\Sigma$, which are not so common in the mapped
region, we have weaker constraints on $P_{\rm 1.3mm}$, but approximate
it as a constant of $\sim 10^{-2}$ in the fiducial cases. Such an
empirical relation can provide a test of theoretical/numerical models
of star formation.

We have then utilized the continuum image and the estimated form of
$P_{\rm 1.3mm}(\Sigma)$ to carry out various estimates of the
``dense'' gas mass fraction, $f_{\rm dg}$, in the IRDC and, by
extrapolation with the observed $\Sigma$-PDF, in the larger-scale GMC
region. The mass estimate in the mapped region of the IRDC made
directly from the observed 1.3~mm flux depends on adopted dust
opacities and temperatures, but has a fiducial value of just under
10\%. Using the MIREX $\Sigma$ at location of 1.3~mm flux detection leads
to mass fraction estimates that are about a factor of 1.5 times
higher. Extrapolating to the larger scale region, given its observed
log-normal $\Sigma$-PDF, we find values of $f_{\rm dg}\sim 7\%$.

Finally, assuming that the detected 1.3~mm structures mostly trace
protostellar cores and capture the bulk of the mass of the core
population, we use these results to estimate the star formation rate
in the IRDC, in particular the star formation efficiency per free-fall
time, $\epsilon_{\rm ff}$. This analysis requires a model to link core
properties to ambient clump properties, for which we utilize the
Turbulent Core Model of \citet{2003ApJ...585..850M}. Then individual star
formation times are, on average, about half of the clump free-fall
time. Given an expected core to star formation efficiency,
$\epsilon_{\rm core}$, of about 50\%, then leads to estimates of
$\epsilon_{\rm ff}\simeq f_{\rm dg} \simeq 10\%$. 

Future improvements in this measurement have been outlined, including
better temperature and thus mass estimates of the protostellar cores
and confirmation of protostellar activity via analysis of outflow
properties. Future work may also include extension of these methods to
a larger sample of IRDCs and star-forming regions.

\acknowledgments 
We thank Alyssa Goodman for constructive suggestions to the paper.
We thank Charles Lada, Richard Larson, Nick Scoville,
Adam Leroy, Gus Oemler, Qizhou Zhang, John Carpenter, and Wanggi Lim for fruitful
discussions. SK was funded by NSF award AST-1140063 while conducting this study.
JCT acknowledges NSF grant AST1411527.  PC acknowledges
the financial support of the European Research Council (ERC; project
PALs 320620).  This paper makes use of the following ALMA data:
ADS/JAO.ALMA\#2015.1.00183.S. ALMA is a partnership of ESO
(representing its member states), NSF (USA) and NINS (Japan), together
with NRC (Canada), NSC and ASIAA (Taiwan), and KASI (Republic of
Korea), in cooperation with the Republic of Chile.  The Joint ALMA
Observatory is operated by ESO, AUI/NRAO and NAOJ.  The National Radio
Astronomy Observatory is a facility of the National Science Foundation
operated under cooperative agreement by Associated Universities, Inc.

{\it Facilities:} \facility{ALMA}; 


\begin{thebibliography}{}
\expandafter\ifx\csname natexlab\endcsname\relax\def\natexlab#1{#1}\fi

\bibitem[{{Alves} {et~al.}(2017){Alves}, {Lombardi}, \&
  {Lada}}]{2017A&A...606L...2A}
{Alves}, J., {Lombardi}, M., \& {Lada}, C.~J. 2017, \aap, 606, L2

\bibitem[{{Andr{\'e}} {et~al.}(2014){Andr{\'e}}, {Di Francesco},
  {Ward-Thompson}, {Inutsuka}, {Pudritz}, \& {Pineda}}]{2014prpl.conf...27A}
{Andr{\'e}}, P., {Di Francesco}, J., {Ward-Thompson}, D., {et~al.} 2014,
  Protostars and Planets VI, 27

\bibitem[{{Beltr{\'a}n} \& {de Wit}(2016)}]{2016A&ARv..24....6B}
{Beltr{\'a}n}, M.~T., \& {de Wit}, W.~J. 2016, \aapr, 24, 6

\bibitem[{{Bergin} \& {Tafalla}(2007)}]{2007ARA&A..45..339B}
{Bergin}, E.~A., \& {Tafalla}, M. 2007, \araa, 45, 339

\bibitem[{{Butler} \& {Tan}(2009)}]{2009ApJ...696..484B}
{Butler}, M.~J., \& {Tan}, J.~C. 2009, \apj, 696, 484

\bibitem[{{Butler} \& {Tan}(2012)}]{2012ApJ...754....5B}
---. 2012, \apj, 754, 5

\bibitem[{{Butler} {et~al.}(2014){Butler}, {Tan}, \&
  {Kainulainen}}]{2014ApJ...782L..30B}
{Butler}, M.~J., {Tan}, J.~C., \& {Kainulainen}, J. 2014, \apjl, 782, L30

\bibitem[{{Chabrier} {et~al.}(2014){Chabrier}, {Hennebelle}, \&
  {Charlot}}]{2014ApJ...796...75C}
{Chabrier}, G., {Hennebelle}, P., \& {Charlot}, S. 2014, \apj, 796, 75

\bibitem[{{Chen} {et~al.}(2017){Chen}, {Burkhart}, {Goodman}, \&
  {Collins}}]{2017arXiv170709356C}
{Chen}, H., {Burkhart}, B., {Goodman}, A.~A., \& {Collins}, D.~C. 2017, ArXiv
  e-prints, arXiv:1707.09356

\bibitem[{{Christie} {et~al.}(2017){Christie}, {Wu}, \&
  {Tan}}]{2017ApJ...848...50C}
{Christie}, D., {Wu}, B., \& {Tan}, J.~C. 2017, \apj, 848, 50

\bibitem[{{Churchwell} {et~al.}(2009){Churchwell}, {Babler}, {Meade},
  {Whitney}, {Benjamin}, {Indebetouw}, {Cyganowski}, {Robitaille}, {Povich},
  {Watson}, \& {Bracker}}]{2009PASP..121..213C}
{Churchwell}, E., {Babler}, B.~L., {Meade}, M.~R., {et~al.} 2009, \pasp, 121,
  213

\bibitem[{{Collins} {et~al.}(2011){Collins}, {Padoan}, {Norman}, \&
  {Xu}}]{2011ApJ...731...59C}
{Collins}, D.~C., {Padoan}, P., {Norman}, M.~L., \& {Xu}, H. 2011, \apj, 731,
  59

\bibitem[{{Da Rio} {et~al.}(2014){Da Rio}, {Tan}, \&
  {Jaehnig}}]{2014ApJ...795...55D}
{Da Rio}, N., {Tan}, J.~C., \& {Jaehnig}, K. 2014, \apj, 795, 55

\bibitem[{{De Buizer} {et~al.}(2017){De Buizer}, {Liu}, {Tan}, {Zhang},
  {Beltr{\'a}n}, {Shuping}, {Staff}, {Tanaka}, \&
  {Whitney}}]{2017ApJ...843...33D}
{De Buizer}, J.~M., {Liu}, M., {Tan}, J.~C., {et~al.} 2017, \apj, 843, 33

\bibitem[{{Draine}(2011)}]{2011piim.book.....D}
{Draine}, B.~T. 2011, {Physics of the Interstellar and Intergalactic Medium},
  Princeton Series in Astrophysics (Princeton University Press)

\bibitem[{{Federrath} \& {Klessen}(2013)}]{2013ApJ...763...51F}
{Federrath}, C., \& {Klessen}, R.~S. 2013, \apj, 763, 51

\bibitem[{{Goodman} {et~al.}(2009){Goodman}, {Pineda}, \&
  {Schnee}}]{2009ApJ...692...91G}
{Goodman}, A.~A., {Pineda}, J.~E., \& {Schnee}, S.~L. 2009, \apj, 692, 91

\bibitem[{{Hennebelle} \& {Chabrier}(2008)}]{2008ApJ...684..395H}
{Hennebelle}, P., \& {Chabrier}, G. 2008, \apj, 684, 395

\bibitem[{{Hennebelle} \& {Chabrier}(2011)}]{2011ApJ...743L..29H}
---. 2011, \apjl, 743, L29

\bibitem[{{Johnstone} {et~al.}(2004){Johnstone}, {Di Francesco}, \&
  {Kirk}}]{2004ApJ...611L..45J}
{Johnstone}, D., {Di Francesco}, J., \& {Kirk}, H. 2004, \apjl, 611, L45

\bibitem[{{Kainulainen} {et~al.}(2009){Kainulainen}, {Beuther}, {Henning}, \&
  {Plume}}]{2009A&A...508L..35K}
{Kainulainen}, J., {Beuther}, H., {Henning}, T., \& {Plume}, R. 2009, \aap,
  508, L35

\bibitem[{{Kainulainen} \& {Tan}(2013)}]{2013A&A...549A..53K}
{Kainulainen}, J., \& {Tan}, J.~C. 2013, \aap, 549, A53

\bibitem[{{Kennicutt} \& {Evans}(2012)}]{2012ARA&A..50..531K}
{Kennicutt}, R.~C., \& {Evans}, N.~J. 2012, \araa, 50, 531

\bibitem[{{Kong} {et~al.}(2017){Kong}, {Tan}, {Caselli}, {Fontani}, {Wang}, \&
  {Butler}}]{2017arXiv170105953K}
{Kong}, S., {Tan}, J.~C., {Caselli}, P., {et~al.} 2017, ArXiv e-prints,
  arXiv:1701.05953

\bibitem[{{Krumholz} {et~al.}(2012){Krumholz}, {Klein}, \&
  {McKee}}]{2012ApJ...754...71K}
{Krumholz}, M.~R., {Klein}, R.~I., \& {McKee}, C.~F. 2012, \apj, 754, 71

\bibitem[{{Krumholz} \& {McKee}(2005)}]{2005ApJ...630..250K}
{Krumholz}, M.~R., \& {McKee}, C.~F. 2005, \apj, 630, 250

\bibitem[{{Krumholz} \& {Tan}(2007)}]{2007ApJ...654..304K}
{Krumholz}, M.~R., \& {Tan}, J.~C. 2007, \apj, 654, 304

\bibitem[{{Kunz} \& {Mouschovias}(2009)}]{2009MNRAS.399L..94K}
{Kunz}, M.~W., \& {Mouschovias}, T.~C. 2009, \mnras, 399, L94

\bibitem[{{Lada} {et~al.}(2010){Lada}, {Lombardi}, \&
  {Alves}}]{2010ApJ...724..687L}
{Lada}, C.~J., {Lombardi}, M., \& {Alves}, J.~F. 2010, \apj, 724, 687

\bibitem[{{Lee} {et~al.}(2016){Lee}, {Miville-Desch{\^e}nes}, \&
  {Murray}}]{2016ApJ...833..229L}
{Lee}, E.~J., {Miville-Desch{\^e}nes}, M.-A., \& {Murray}, N.~W. 2016, \apj,
  833, 229

\bibitem[{{Li} \& {Shu}(1997)}]{1997ApJ...475..237L}
{Li}, Z.-Y., \& {Shu}, F.~H. 1997, \apj, 475, 237

\bibitem[{{Lim} {et~al.}(2016){Lim}, {Tan}, {Kainulainen}, {Ma}, \&
  {Butler}}]{2016ApJ...829L..19L}
{Lim}, W., {Tan}, J.~C., {Kainulainen}, J., {Ma}, B., \& {Butler}, M.~J. 2016,
  \apjl, 829, L19

\bibitem[{{Lombardi}(2009)}]{2009A&A...493..735L}
{Lombardi}, M. 2009, \aap, 493, 735

\bibitem[{{Matsushita} {et~al.}(2017){Matsushita}, {Machida}, {Sakurai}, \&
  {Hosokawa}}]{2017MNRAS.470.1026M}
{Matsushita}, Y., {Machida}, M.~N., {Sakurai}, Y., \& {Hosokawa}, T. 2017,
  \mnras, 470, 1026

\bibitem[{{Matzner} \& {McKee}(2000)}]{2000ApJ...545..364M}
{Matzner}, C.~D., \& {McKee}, C.~F. 2000, \apj, 545, 364

\bibitem[{{McKee}(1989)}]{1989ApJ...345..782M}
{McKee}, C.~F. 1989, \apj, 345, 782

\bibitem[{{McKee} \& {Tan}(2003)}]{2003ApJ...585..850M}
{McKee}, C.~F., \& {Tan}, J.~C. 2003, \apj, 585, 850

\bibitem[{{Murray}(2011)}]{2011ApJ...729..133M}
{Murray}, N. 2011, \apj, 729, 133

\bibitem[{{Myers}(2015)}]{2015ApJ...806..226M}
{Myers}, P.~C. 2015, \apj, 806, 226

\bibitem[{{Offner} {et~al.}(2014){Offner}, {Clark}, {Hennebelle}, {Bastian},
  {Bate}, {Hopkins}, {Moraux}, \& {Whitworth}}]{2014prpl.conf...53O}
{Offner}, S.~S.~R., {Clark}, P.~C., {Hennebelle}, P., {et~al.} 2014, Protostars
  and Planets VI, 53

\bibitem[{{Ossenkopf} \& {Henning}(1994)}]{1994A&A...291..943O}
{Ossenkopf}, V., \& {Henning}, T. 1994, \aap, 291, 943

\bibitem[{{Padoan} \& {Nordlund}(2002)}]{2002ApJ...576..870P}
{Padoan}, P., \& {Nordlund}, {\AA}. 2002, \apj, 576, 870

\bibitem[{{Padoan} \& {Nordlund}(2011)}]{2011ApJ...741L..22P}
---. 2011, \apjl, 741, L22

\bibitem[{{Sanhueza} {et~al.}(2017){Sanhueza}, {Jackson}, {Zhang},
  {Guzm{\'a}n}, {Lu}, {Stephens}, {Wang}, \& {Tatematsu}}]{2017ApJ...841...97S}
{Sanhueza}, P., {Jackson}, J.~M., {Zhang}, Q., {et~al.} 2017, \apj, 841, 97

\bibitem[{{Scoville} {et~al.}(1986){Scoville}, {Sanders}, \&
  {Clemens}}]{1986ApJ...310L..77S}
{Scoville}, N.~Z., {Sanders}, D.~B., \& {Clemens}, D.~P. 1986, \apjl, 310, L77

\bibitem[{{Shu} {et~al.}(1987){Shu}, {Adams}, \&
  {Lizano}}]{1987ARA&A..25...23S}
{Shu}, F.~H., {Adams}, F.~C., \& {Lizano}, S. 1987, \araa, 25, 23

\bibitem[{{Stutz} \& {Kainulainen}(2015)}]{2015A&A...577L...6S}
{Stutz}, A.~M., \& {Kainulainen}, J. 2015, \aap, 577, L6

\bibitem[{{Tan}(2000)}]{2000ApJ...536..173T}
{Tan}, J.~C. 2000, \apj, 536, 173

\bibitem[{{Tan}(2016)}]{2016IAUS..315..154T}
{Tan}, J.~C. 2016, in IAU Symposium, Vol. 315, From Interstellar Clouds to
  Star-Forming Galaxies: Universal Processes?, ed. P.~{Jablonka},
  P.~{Andr{\'e}}, \& F.~{van der Tak}, 154--162

\bibitem[{{Tan} {et~al.}(2014){Tan}, {Beltr{\'a}n}, {Caselli}, {Fontani},
  {Fuente}, {Krumholz}, {McKee}, \& {Stolte}}]{2014prpl.conf..149T}
{Tan}, J.~C., {Beltr{\'a}n}, M.~T., {Caselli}, P., {et~al.} 2014, Protostars
  and Planets VI, 149

\bibitem[{{Tan} {et~al.}(2013){Tan}, {Kong}, {Butler}, {Caselli}, \&
  {Fontani}}]{2013ApJ...779...96T}
{Tan}, J.~C., {Kong}, S., {Butler}, M.~J., {Caselli}, P., \& {Fontani}, F.
  2013, \apj, 779, 96

\bibitem[{{Tan} {et~al.}(2016){Tan}, {Kong}, {Zhang}, {Fontani}, {Caselli}, \&
  {Butler}}]{2016ApJ...821L...3T}
{Tan}, J.~C., {Kong}, S., {Zhang}, Y., {et~al.} 2016, \apjl, 821, L3

\bibitem[{{Tanaka} {et~al.}(2017){Tanaka}, {Tan}, \&
  {Zhang}}]{2017ApJ...835...32T}
{Tanaka}, K.~E.~I., {Tan}, J.~C., \& {Zhang}, Y. 2017, \apj, 835, 32

\bibitem[{{Wu} {et~al.}(2017){Wu}, {Tan}, {Christie}, {Nakamura}, {Van Loo}, \&
  {Collins}}]{2017ApJ...841...88W}
{Wu}, B., {Tan}, J.~C., {Christie}, D., {et~al.} 2017, \apj, 841, 88

\bibitem[{{Wu} {et~al.}(2015){Wu}, {Van Loo}, {Tan}, \&
  {Bruderer}}]{2015ApJ...811...56W}
{Wu}, B., {Van Loo}, S., {Tan}, J.~C., \& {Bruderer}, S. 2015, \apj, 811, 56

\bibitem[{{Zhang} {et~al.}(2015){Zhang}, {Wang}, {Lu}, \&
  {Jim{\'e}nez-Serra}}]{2015ApJ...804..141Z}
{Zhang}, Q., {Wang}, K., {Lu}, X., \& {Jim{\'e}nez-Serra}, I. 2015, \apj, 804,
  141

\bibitem[{{Zhang} {et~al.}(2009){Zhang}, {Wang}, {Pillai}, \&
  {Rathborne}}]{2009ApJ...696..268Z}
{Zhang}, Q., {Wang}, Y., {Pillai}, T., \& {Rathborne}, J. 2009, \apj, 696, 268

\bibitem[{{Zhang} \& {Tan}(2015)}]{2015ApJ...802L..15Z}
{Zhang}, Y., \& {Tan}, J.~C. 2015, \apjl, 802, L15

\bibitem[{{Zhang} \& {Tan}(2017)}]{2017arXiv170808853Z}
---. 2017, ArXiv e-prints, arXiv:1708.08853

\bibitem[{{Zhang} {et~al.}(2014){Zhang}, {Tan}, \&
  {Hosokawa}}]{2014ApJ...788..166Z}
{Zhang}, Y., {Tan}, J.~C., \& {Hosokawa}, T. 2014, \apj, 788, 166

\bibitem[{{Zuckerman} \& {Evans}(1974)}]{1974ApJ...192L.149Z}
{Zuckerman}, B., \& {Evans}, II, N.~J. 1974, \apjl, 192, L149

\end{thebibliography}

\end{document}